\documentclass[review,onefignum,onetabnum]{siamonline171218}

\usepackage{lipsum}
\usepackage{amsfonts}
\usepackage{amssymb}
\usepackage{bigints}
\usepackage{graphicx}
\usepackage{epstopdf}
\usepackage{algorithmic}
\usepackage{enumitem}
\usepackage{booktabs}
\usepackage{multirow}
\ifpdf
  \DeclareGraphicsExtensions{.eps,.pdf,.png,.jpg}
\else
  \DeclareGraphicsExtensions{.eps}
\fi

\newsiamremark{remark}{Remark}
\newsiamremark{hypothesis}{Hypothesis}
\crefname{hypothesis}{Hypothesis}{Hypotheses}
\newsiamthm{claim}{Claim}

\headers{a competitive respiratory disease system}{A.D Fome, and W. Bock, and A. Klar}

\title{Analysis of a competitive respiratory disease system with quarantine\thanks{Submitted to the editors DATE.
\funding{This work was funded by ....}}}

\author{Anna Daniel Fome\footnotemark[2]\ \footnotemark[3]\thanks{Rheinland-Pfälzische Technische Universität Kaiserslautern-Landau, 67663 Kaiserslautern, Germany 
  (\email{fome@mathematik.uni-kl.de}, \email{bock@mathematik.uni-kl.de}, \email{klar@mathematik.uni-kl.de}).} \and Wolfgang Bock\footnotemark[2], 
\and Axel Klar\footnotemark[2]}

\footnotetext[3]{Jordan university: a constituent college of St. Augustine University of Tanzania, Morogoro, Tanzania. }

\usepackage{amsopn}

\makeatletter
\newcommand*{\addFileDependency}[1]{
  \typeout{(#1)}
  \@addtofilelist{#1}
  \IfFileExists{#1}{}{\typeout{No file #1.}}
}
\makeatother

\newcommand*{\myexternaldocument}[1]{%
    \externaldocument{#1}%
    \addFileDependency{#1.tex}%
    \addFileDependency{#1.aux}%
}

\ifpdf
\hypersetup{
  pdftitle={A COMPETITIVE RESPIRATORY DISEASE SYSTEM},
  pdfauthor={A.D. Fome, W. Bock, and A. Klar}
}
\fi

\myexternaldocument{ex_supplement}

\begin{document}
\nolinenumbers

\maketitle
\begin{abstract}
   In the world of epidemics, the mathematical modeling of disease co-infection is gaining importance due to its contributions to mathematics and public health. Because the co-infection may have a double burden on families, countries, and the universe, understanding its dynamics is paramount. We study a SEIQR (susceptible-exposed-infectious-quarantined-recovered) deterministic epidemic model with a single host population and multiple strains (-$c$ and -$i$) to account for two competitive diseases with quarantine effects. To model the role of quarantine and isolation efficacy in disease dynamics, we utilize a linear function. Further, we shed light on the standard endemic threshold and determine the conditions for extinction or coexistence with and without forming co-infection. Next, we show the dependence of the criticality based on specific parameters of the different pathogens. We found that the disease-free equilibrium (DFE) of the single-strain model always exists and is globally asymptotically stable (GAS) if $\tilde{\mathcal{R}}_k^q\leq 1$, else, a stable endemic equilibrium. On top of that, the model has forward bifurcation at $\tilde{\mathcal{R}}_k^q = 1$. In the case of a two-strain model, the strain with a large reproduction number outcompetes the one with a smaller reproduction number. Further, if the co-infected quarantine reproduction number is less than one, the infections of already infected individuals will die out, and co-infection will persist in the population otherwise. We note that the quarantine and isolation of exposed and infected individuals will reduce the number of secondary cases below one, consequently reducing the disease complications if the total number of people in the quarantine is at most the critical value.

\end{abstract}



\section{Introduction}
		
		An acute respiratory disease (ARI)  is amongst the five respiratory conditions responsible for the tremendous burden to the society \cite{marciniuk2014respiratory},  particularly in low- and middle-income countries \cite{cruz2007global, world2014infection}. It is a condition caused by an infectious agent, viruses, or mixed viral-bacterial infections that disturb body organs and the airway system. Some of the pathogens that cause this condition include influenza virus as well as severe acute respiratory syndrome coronavirus (SARS CoV) \cite{world2014infection} of which they are responsible for several pandemics and emergencies of global consideration; for example, the most severe flu pandemic known as \textit{Spanish flu} which occurred in 1918. The outbreak took approximately 50 million lives and infected half of the globe's population \cite{kitler2002influenza}. 
	
	Since 1918,  the human worries and great suffering of individuals, families, and communities changed the world's impression of flu-like illness \cite{kitler2002influenza}. In recent years, the emergence of severe acute respiratory syndrome coronavirus 2 (SARS-CoV-2) has demonstrated beyond doubt its ability to infect humans and cause widespread outbreaks. The virus originated in Wuhan City, China, at the beginning of December 2019 (henceforth Covid-19). A few months later (On March 11. 2020), the World Health Organization (WHO) declared Covid-19 as a global outbreak of pandemic\cite{jebril2020world}. 

    The increased number of reported infections and deaths due to the flu-like pandemic inspired many researchers to investigate the co-exists of multiple illnesses. Several clinical studies on the subject are particularly relevant \cite{alosaimi2021influenza, dadashi2021covid, konala2020co, lai2020co, lansbury2020co,  singh2020covid, stefanska2013co, zhu2020co}. To manage coexistence, distinguishing the two conditions from each other is necessary. However, due to similar clinical features of the existing diseases, identifying the etiologic agent through laboratory testing is recommended \cite{lai2020co}. Moreover, as flu-like infections (in particular coronavirus and influenza) prompt severe complications and high mortality in a population, the co-infections suggest the consideration of comorbidity \cite{alosaimi2021influenza}.
	 
	 	Furthermore, bio-mathematicians and medical experts are considerable regarding acute disease management \cite{agarwal2021analysis} to understand the dynamics of some co-infection regardless of the novelty of the disease and the complexity of the profound etiology. The assimilation of these interactions through mathematical models to inform the population regarding the disease condition and the uncertainty linked with the infection have been studied \cite{acuna2021co, bhowmick2023decoding, fudolig2020local,  khyar2020global, lazebnik2022generic, ojo2022nonlinear, perez2022model, pinky2020sars, soni2021covid}  and the references therein.  Based on control measures: mathematical modeling studies have depicted a positive impact upon applying quarantine during the disease outbreak \cite{ali2020role, aronna2021model, ashcroft2021quantifying, chladna2021effect}.
   
		We introduce a deterministic $SEIQR$-type epidemic system to account for two competitive strains: a typical example of such pathogens can be influenza (strain-$i$) and SARS-Cov-2 (strain-$c$).  We include a quarantine class with the idea that quarantine and isolation might affect the disease's dynamic.  Additionally, we shed light on the standard endemic threshold and show the dependence of the criticality based on specific parameters of the different pathogens. 

       The rest of the paper is as follows: the details of model assumptions, formulation, and basic properties are in the subsequent section. We analyzed the stability of disease-free and endemic equilibria and the bifurcation analysis of a single-strain model with imperfect quarantine in Section 3. In addition to that,  we presented multiple-strain models (MSM)  with and without co-infected individuals, sensitivity, and the impact of quarantine. Lastly, Section 4 detailed the conclusions, remarks, and future work.

	\section{Model framework}\label{sec:model_framework}
	We assumed that the population is well-mixed and the chance to contact one another is the same. The total population size denoted by $N_T(t)$ split into thirteen compartments: the individuals who are vulnerable to getting a disease upon contact with infectious denoted by $S$; the exposed individuals are $E_i$, $E_c$, and $E_{ic}$; individuals who are capable spreading the disease are $I_i$, $I_c$, and $I_{ic}$; the classes $Q_i$, $Q_c$, and $Q_{ic}$ are the individuals quarantined or detached; as well as  $R_i$, $R_c$, and $R_{ic}$,  the recovered individuals. 
 
\subsection{Main assumption statements}\label{ssst:Model assmp}
	\begin{enumerate}
		\item \textit{\textbf{Vital dynamics:}}  In 1989,   in his literature  \cite{hethcote1989three}, Hethcote argued the fact of including inflow (birth) and outflow (death)  when modeling the disease persisted in a population; so we assumed our model to reflect such kind of diseases.
        
        \item \textit{\textbf{Formation of co-infected class}:} If an individual carries two or more pathogens, that person is co-infected. Therefore, we assume that when a symptomatic individual with pathogen $i$ gets in contact with another individual infected with pathogen $c$, or vice-versa the co-exposed class  $E_{ic}$ formed. On the contrary, co-existence is the state or fact of both strains existing in the same population at time $t$.

		\item\textbf{\textit{Quarantine and Isolation: }} We assumed that quarantined individuals are either recognized as a recent contact with a confirmed case or symptomatic. We have ignored the quarantine of the healthy population. To be clear and specific, we define an imperfect quarantine: as any quarantine site or facility (including home) characterized by violating any guidelines, rules, and regulations to control the spread of infections. For instance
		\begin{enumerate}
		\item  \textit{Home quarantine}: an individual may want to go out for several reasons, like shopping or getting fresh air before quarantine maturity time. In such a case,  if the individual is carrying the disease can transmit the virus (however, at a reduced rate). 
		
        \item\textit{Campsite quarantine}: In this case, an exposed individual (not carrying the virus) may become infected in the campsite due to the movements or activities when the quarantine rules are violated. 
        \end{enumerate}

		We employ a linear function \cref{eqn:DecreasingFunction} studied by \cite{fome2023deterministic, mitchell2016data}  to portray a reduction of disease transmission because of introducing control measures. We used the parameter $\omega_k$ to measure the potential inputs provided by the respective authority and assume it is proportional to the number of infected individuals. The inputs include but are not limited to biomedical waste management, provision of medical equipment,  social and psychosocial support, and protections, as well as support to meet their basic needs. Moreover, education about infection preventive measures and the importance of promptly seeking medical care if they develop symptoms \cite{world2020considerations}. We let $\Omega_k(I_k)=\Omega_{k}$ be a continuously differentiable function w.r.t. to $I_k$  defined as
  
		\begin{align}
			\Omega_k=1-\omega_kI_k=
			\begin{cases}
				0, & \text{ if } ~~I_k\omega_k=1,\\
				(0,1], & \text{ if }~~ I_k\omega_k\in [0,1),\\
			\end{cases}
			\label{eqn:DecreasingFunction}
		\end{align}
where $k$ stands for subscript $i,c,ic$. For all $I_k\in(0,1)$ we assume $\omega_k\in [0,1]$ (hence $\Omega_k\notin \mathbb{R}^-$). As an approximation to function \cref{eqn:DecreasingFunction}, $\omega_kI_k$ and $\Omega_k$ satisfies the following as $\omega_k$ and $I_k$ varies:
 \begin{itemize}
 \item[] \textit{Perfect quarantine:} Note, we assumed that, $\omega_k$ is proportional to $I_k$: hence as $I_k\to 1$, also $\omega_k\to 1$, and $I_k\omega_k\to 1$. For perfect quarantine, we hold that $I_k\omega_k\approx 1$ so, $\Omega_k=0$. 

  \item[] \textit{Imperfect quarantine:} 
  In this case, for all $I_{k} \in(0,1)$ and   $\omega_{k} \in(0,1]$ we have the following estimates: 
 \end{itemize}
   \begin{table}[ht]
   \centering
	\begin{tabular}{lllc}
		\hline
		\multicolumn{2}{c}{\textbf{Changes in}} & &\multirow{2}{*}{\textbf{Level of imperfection}}\\\cline{1-2}
  \textbf{$\omega_k$, $I_k$} & $\omega_kI_k$, $\Omega_k$ & & \\	\hline
  $\omega_k\to 0, I_k\to 0$ &   $\omega_kI_k\to 0 $, $\Omega_k\to 1$ && high\\
   $\omega_k\to 0, I_k\to 1$ &   $\omega_kI_k\to 0 $, $\Omega_k\to 1$ & &high\\
    $\omega_k\to 1, I_k\to 0$ &   $\omega_kI_k\to 0 $, $\Omega_k\to 1$ && high\\
     $\omega_k\to 1, I_k\to 1$ &   $\omega_kI_k\to 1 $, $\Omega_k\to 0$ && low\\\hline
	\end{tabular}
	\caption{Estimated level of imperfection}
	\label{tab:ImperfectLevel}
\end{table}

  From equation \cref{eqn:DecreasingFunction} and \cref{tab:ImperfectLevel}, one can observe that "the smaller the value of $\Omega_k$ is, the effective quarantine is; in other words, as the value of $\Omega_{k}$ approaches zero, the smaller the chance of quarantined individuals to be infected. 
  
		\item \textit{\textbf{Cross-immunity:}}  A recent study by Almazán and her colleagues \cite{almazan2021influenza}  investigated how influenza's pre-existing immunity to SARS-CoV-2 predicts Covid-19 dynamics. The study identified eleven CD8 T-cell peptides cross-reacted with flu and SARS-CoV-2 pathogens depending on the leukocyte antigen type. The detailed mathematical model of cross-immunity is given briefly in the book by Martcheva \cite{martcheva2015introduction} and detailed further in \cite{castillo1989epidemiological, nuno2005dynamics}. Thus, with parameters $\eta_i,\eta_c\in[0,1]$, we model the immune response for individuals who have recovered from  $pathogen~i$ to be sick with $pathgen~c$ or contrariwise.  The values $\eta_i=0$ and $\eta_c=0$ correspond to full cross-immunity, and $\eta_i=1$ and $\eta_c=1$ equals no cross-immunity. We further assume that an individual who recovered from one strain has permanent immunity to that strain. 		
\end{enumerate}

	\begin{figure}[ht]
		\centering
	\includegraphics[width=1.\linewidth]{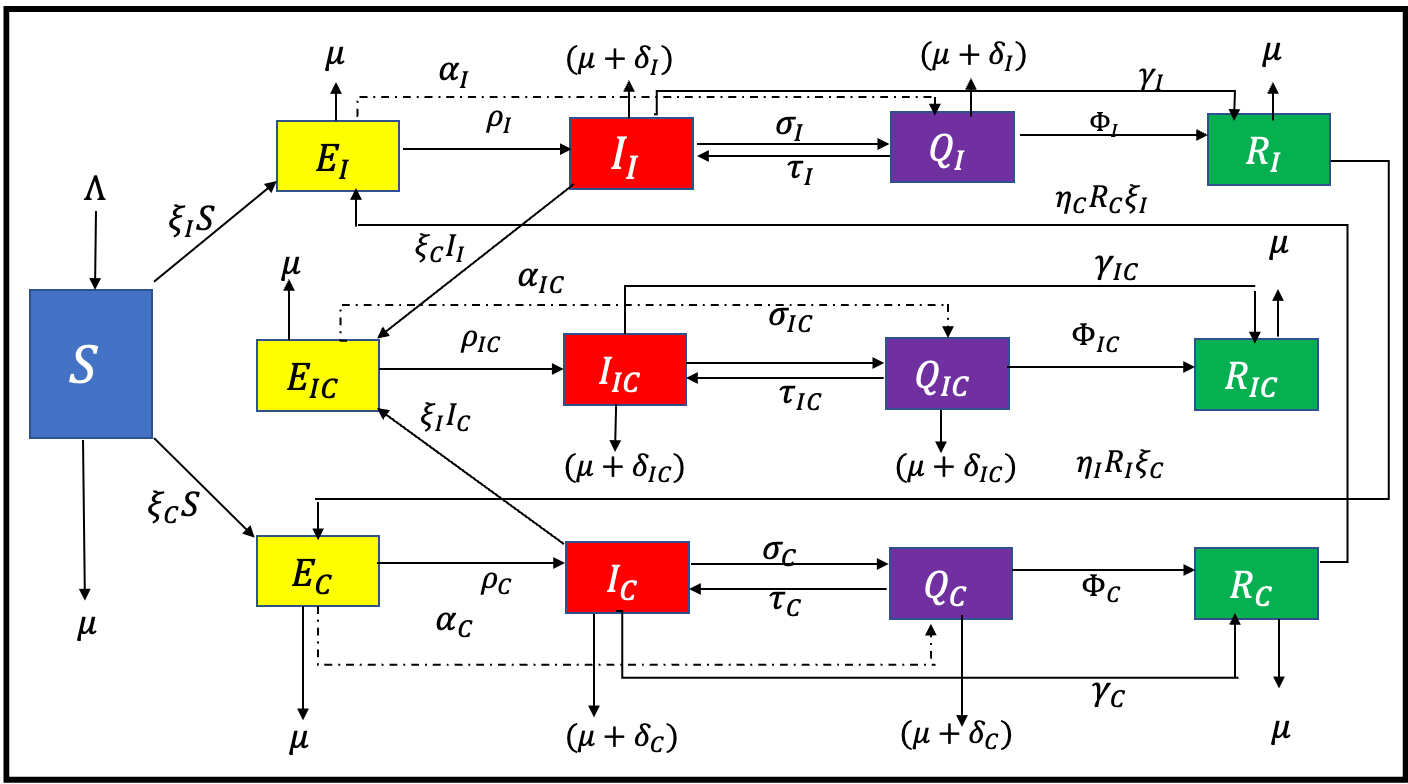}
		\caption{Two-strain epidemic with cross-immunity and  co-infection}
        \label{fig:Mainmodel}
	\end{figure}
	\subsection{Model formulation}\label{ssst:Model forml}
We assumed that all individuals are recruited into the susceptible population $S$ with a constant inflow rate $\Lambda$ and decrease by the rate $\mu$, natural death from each class. The susceptible individuals become infected via contact with an infectious individual and move into the exposed class $E_i$ or $E_c$. At the rates $\rho_i$ and $\rho_{c}$ the portions of $E_{i}$ and $E_{c}$ will progress to the infectious classes $I_{i}$ and $I_{c}$,  respectively.  The progression of co-exposed individuals to active co-infections occurs at the rate $\rho_{ic}$. With the rates $\alpha_i$, $\alpha_c$ and $\alpha_{ic}$ the exposed individual will enter the $Q_i$, $Q_c$, and $Q_{ic}$ classes and further  to $I_i$, $I_c$, and $I_{ic}$ at the rates $\tau_i$, $\tau_c$ and $\tau_{ic}$, respectively. The growth in the quarantine from the infectious classes is at the rates $\sigma_{i }$, $\sigma_{c }$, and $\sigma_{ic }$. The fractions of infected and quarantine classes will progress to the recovered classes at the recovery rates $\gamma$ and $\phi$  comparable to the number of people in their compartment. We summarize \cref{ssst:Model assmp}, and \cref{ssst:Model forml}  in \cref{fig:Mainmodel}. From the figure, 	$\xi_i$ and  $	\xi_c$ define strain-$i$ and strain-$c$  forces of infections,  respectively, are given by
\begin{equation}
	\xi_i=\dfrac{\beta_i(I_i+I_{ic}+\Omega_{i}Q_i+\Omega_{ic}Q_{ic})}{N_T},  \text{~~~~~and~~~~~}  \xi_c=\dfrac{\beta_c(I_c+I_{ic}+\Omega_{c}Q_c+\Omega_{ic}Q_{ic})}{N_T}. \label{eqn:ForceInf}
\end{equation}
In \cref{eqn:ForceInf}, $\beta_i$ and $\beta_c$ are  the contact rates where $\Omega_{i}$, $\Omega_{c}$  and $\Omega_{ic}$ are the decreasing functions of infectiousness given in \cref{eqn:DecreasingFunction}. Using the above assumptions,  \cref{fig:Mainmodel}, and parameters descriptions in  \cref{tab:ModelParameters}, the following system equations are obtained:	

	\begin{equation}
			\begin{aligned}
					S'(t)=&\Lambda - \xi_{i}S-\xi_{c}S - \mu S,\\
					E'_{i}(t)=& \xi_{i}S+\eta_cR_c\xi_i - \mathbf{p}_iE_i,\\
					E'_{c}(t)=& \xi_{c}S +\eta_iR_i\xi_c- \mathbf{p}_cE_c,\\
					E'_{ic}(t)=& \xi_{i}I_c +\xi_cI_i - \mathbf{p}_{ic}E_{ic},\\
					I'_i(t)=&\rho_iE_i+\tau_iQ_i-\xi_cI_i-\mathbf{q}_iI_i,\\
					I'_c(t)=&\rho_cE_c+\tau_cQ_c-\xi_iI_c-\mathbf{q}_cI_c,\\
					I'_{ic}(t)=&\rho_{ic}E_{ic}+\tau_{ic}Q_{ic}-\mathbf{q}_{ic}I_{ic},\\
					Q'_i(t)=&\alpha_iE_i-\mathbf{r}_iQ_i+\sigma_iI_i,\\
					Q'_c(t)=&\alpha_cE_c-\mathbf{r}_cQ_c+\sigma_cI_c,\\	
					Q'_{ic}(t)=&\alpha_{ic}E_{ic}-\mathbf{r}_{ic}Q_{ic}+\sigma_{ic}I_{ic},\\	
					R'_{i}(t)=&\phi_iQ_i+\gamma_iI_i-\eta_iR_i\xi_c-\mu R_i,\\
					R'_{c}(t)=&\phi_cQ_c+\gamma_cI_c-\eta_cR_c\xi_i-\mu R_c,\\
					R'_{ic}(t)=&\phi_{ic}Q_{ic}+\gamma_{ic}I_{ic}-\mu R_{ic},
				\end{aligned}
			\label{eqn:MainModel}
		\end{equation}
where 
\begin{align*}
	&	\mathbf{p}_i=\alpha_i +\rho_i +\mu,  &	\mathbf{p}_c=\alpha_c  +\rho_c +\mu, ~~~~~~~~& 	\mathbf{p}_{ic}=\alpha_{ic} +\rho_{ic} +\mu, \\
	&	\mathbf{q}_i=\gamma_{i}+\sigma_{i}+\delta_{i}+\mu,  &	\mathbf{q}_c=\gamma_{c}+\sigma_{c}+\delta_{c}+\mu,  ~~~~~~~~& 	\mathbf{q}_{ic}=\gamma_{ic}+\sigma_{ic}+\delta_{ic}+\mu, \\
	&	\mathbf{r}_i=\phi_{i}+\tau_{i}+\delta_{i}+\mu,  &	\mathbf{r}_c=\phi_{c}+\tau_{c}+\delta_{c}+\mu,  ~~~~~~~~& 	\mathbf{r}_{ic}=\phi_{ic}+\tau_{ic}+\delta_{ic}+\mu.
\end{align*}

At a given time $t$,  the total population size $	N_T$ is given by 
\begin{align*}
	N_T(t)=& S(t) + E_i(t) + E_c(t) +E_{ic}(t) + I_i(t) + I_c(t) + I_{ic}(t) \\& + Q_i(t) +Q_c(t) +Q_{ic}(t) + R_i(t) + R_c(t) +R_{ic}(t).
\end{align*}

Moreover, the system \cref{eqn:MainModel} is equipped with strictly non-negative initial data 
\begin{equation}
	\begin{aligned}
		&S(0)\geq 0,~E_i(0)\geq 0, ~ E_c(0)\geq 0, ~E_{ic}(0)\geq 0,\\&I_i(0)\geq 0, ~I_c(0)\geq 0,~ I_{ic}(0)\geq 0, ~Q_i(0)\geq 0,~ Q_c(0)\geq 0, \\&Q_{ic}(0)\geq 0, ~R_i(0)\geq 0, ~R_c(0)\geq 0,  \text{ and } R_{ic}(0)\geq 0.
	\end{aligned}
	\label{eqn:MainModel_InitialCon}
\end{equation}

\begin{table}[htbp]
	\begin{tabular}{llll}
		\hline
		\textbf{Symbol}	&  \textbf{Intepretation}& \textbf{Range/Value} &  \textbf{Source }\\	\hline
		$\Lambda$ &  Recruitment rate &  [0.001, 0.1] & Assumed   \\
		$\beta_i$	& strain-$i$ transmission rate &   [0.5, 2]  &  \cite{nuno2005dynamics}\\
		$\beta_c$	&  strain-$c$ transmission rate & [0.5, 2]  &  \cite{aronna2021model, eikenberry2020mask} \\
		$\omega_i$	& The parameter to measure the potential input in $Q_i$  & $[0,1]$  & Assumed  \\
		$\omega_c$	& The parameter to measure the potential input in $Q_c$& $[0,1]$ &  Assumed \\
		$\omega_{ic}$	&The parameter to measure the potential input in $Q_{ic}$ &$[0,1]$  &  Assumed \\
		$\Omega_i$	& Function to reduce the infection and measure the $Q_i$ efficacy  & $[0,1]$  &  \cite{aronna2021model}\\
		$\Omega_c$	& Function to reduce the infection  and measure the $Q_c$ efficacy & $[0,1]$  & \cite{aronna2021model} \\
		$\Omega_{ic}$	& Function to reduce the infection  and measure the $Q_{ic}$ efficacy  & $[0,1]$  & \cite{aronna2021model} \\
		$\rho_i$	&Rate of developing strain-$i$ symptoms  &  [1/14,1/3] & \cite{eikenberry2020mask}  \\
		$\rho_c$	& Rate of developing strain-$c$ symptoms &  [1/14,1/3] & \cite{eikenberry2020mask}  \\
		$\rho_{ic}$	& Rate of developing co-infections symptoms &   [1/14,1/3] & Assumed \\
		$\alpha_{i}$	&  Quarantine rate for strain-$i$ exposed individuals & [0,2] & Assumed \\
		$\alpha_{c}$	& Quarantine rate for strain-$c$ exposed individuals& [0,2] & Assumed \\
		$\alpha_{ic}$	&  Quarantine rate for co-exposed individuals&  [0,2] & Assumed \\
		$\sigma_{i}$	& Isolation rate for strain-$i$ infected individuals &  [0,2] & Assumed \\
		$\sigma_{c}$	& Isolation rate for strain-$c$ infected individuals & [0,2] & Assumed \\
		$\sigma_{ic}$	&  Isolation rate for co-infected individuals &  [0,2] & Assumed \\
		$\tau_{i}$	& Rate of strain-$i$ quarantine-exposed individuals become $I_{i}$ & $\rho_{i}\times0.15$  & Assumed \\
		$\tau_{c}$	& Rate of strain-$c$ quarantine-exposed individuals become $I_{c}$  &  $\rho_{c}\times0.15$  & Assumed \\
		$\tau_{ic}$	&  Rate of strain-$ic$ quarantine-exposed individuals become $I_{ic}$&  $\rho_{ic}\times0.15$  & Assumed \\
		$\phi_{i}$	& Recovery rate of  strain-$i$ quarantined individuals  &  [0.08,0.14] &\cite{tang2020updated} \\
		$\phi_{c}$	&  Recovery rate of  strain-$c$ quarantined individuals & [0.08,0.14] &\cite{tang2020updated} \\
		$\phi_{ic}$	& Recovery rate of  strain-$ic$  quarantined individuals &  [0.08,0.14] &Assumed \\
		$\gamma_{i}$	& Recovery rate of strain-$i$ infected  individuals & [0.28,0.38] & \cite{tang2020updated} \\
		$\gamma_{c}$	& Recovery rate  of strain-$c$ infected  individuals & [0.28,0.38] & \cite{tang2020updated} \\
		$\gamma_{ic}$	& Recovery rate of co-infected  individuals &  [0.28,0.38] & Assumed \\
		$\eta_{i}$	& Strain-$i$ immunity rate & $[0,1]$ & \cite{nuno2005dynamics} \\
		$\eta_{c}$	&  Strain-$c$ immunity rate & $[0,1]$ & \cite{nuno2005dynamics} \\
		$\delta_{i}$	&Death rate due to strain-$i$  & [0.001,0.1] & \cite{eikenberry2020mask} \\
		$\delta_{c}$	& Death rate due to strain-$c$  & [0.001,0.1] &   \cite{eikenberry2020mask}\\
		$\delta_{ic}$	& Death rate due to co-infection & [0.001,0.1] &  Assumed \\
		$\mu$	& natural death rate  & [0.001, 0.1] & Assumed \\\\
		\hline
	\end{tabular}
	\caption{Model Parameters}
	\label{tab:ModelParameters}
\end{table}

\subsection{Basic Properties of the main model}\label{ssst:basic_properties}
	Because we are dealing with problems related to population dynamics, for meaningful biological interpretation, all the variables must be positive and bounded for all time $t\geq 0$. Proofs are in the \cref{lem:Postivity_Boundedness}.
 

\section{Results and discussion}
\subsection{Analysis of Single-Strain model-SSM}\label{ssst:SSM_Analysis}
Before analyzing the entire model, it is significant to investigate the dynamics of the single-strain model (SSM), this means that  either strain-$c$ or strain-$i$ is absent. To generalize, we ask $I_k$ to be some infected individuals with  strain-$i$ or strain-$c$. Here, we assume that the subscript $k$ stands for either $i$ or $c$. To this extent, the reduced imperfect model ( i.e., $\Omega_k\in(0,1]$) has eight equations less than model \cref{eqn:MainModel}:
 
	\begin{equation}
		\begin{aligned}
			S'(t)=&\Lambda -\xi_{k}S - \mu S,\\
			E'_{k}(t)=& \xi_{k}S -\mathbf{p}_kE_k,\\
			I'_{k}(t)=&\rho_kE_k+\tau_kQ_k-\mathbf{q}_kI_k\\
			Q'_{k}(t)=&\alpha_kE_k-\mathbf{r}_kQ_k+\sigma_kI_k,\\	
			R'_{k}(t)=&\phi_kQ_k+\gamma_kI_k-\mu R_k,\\
			\label{eqn:MainModel_Reduced}
		\end{aligned}
	\end{equation}
 	\noindent where the force of infection and the total population size of the reduced model are correspondingly given by
	\begin{equation}
	\begin{array}{lll}
		\xi_k=\dfrac{\beta_k(I_k+\Omega_{k}Q_k)}{N_k }, & 	\text{ and, } &N'_k(t) = \Lambda - \mu N_k - \delta_k (I_k + Q_k).\\
		\label{eqn:ForceInf_Imperfect_Reduced}
	\end{array}
	\end{equation} 
We suspend the proofs of positivity and boundness of the solutions in \cref{app:Postivity_Boundedness_Model_Reduced}.
	
	\subsubsection{Disease-free equilibrium (DFE)}\label{ssst:SSM_DFE} The DFE of model \cref{eqn:MainModel_Reduced}  is given by
		\begin{equation}
	\begin{aligned}
	\mathbf{E}^0_k & = (S_k^0, E_k^0, I_{k}^0, Q_{k}^0,R_k^0 ) = \left(\dfrac{\Lambda}{\mu},0,0,0,0\right).
				\label{eqn:DFE_ReducedModel}
			\end{aligned}
		\end{equation}
	 It is well-known that the local stability of the DFE is associated with the disease reproduction number. Thus, to characterize the reproduction number corresponding to \cref{eqn:MainModel}, we have to look at the local stability of \cref{eqn:DFE_ReducedModel}. Using notations in \cite{van2002reproduction}, we linearise the  positive candidates of the infection terms, $F$, and  a non-singular $M$-matrix, $V$, to obtain
	 \begin{equation*}
	 \begin{array}{llll}
	 F=\begin{pmatrix}
	 	0 &\dfrac{S^0_k}{N_k^0}\beta_k&\dfrac{S^0_k}{N_k^0}\beta_k\\ 
	 	0 &0&0\\
	 	 0 &0&0\\
	 \end{pmatrix} & \text{ and } & 	 V=\begin{pmatrix}
	 \mathbf{p}_k &0&0\\ 
	 -\rho_k & \mathbf{q}_k&-\tau_k\\
	 -\alpha_{k} &-\sigma_{k}& \mathbf{r}_k\\
 \end{pmatrix}, & \text{respectively.}
	 \end{array}
	 \end{equation*}
	 
\noindent It follows from \cite{castillo2002computation, van2002reproduction} that the maximum absolute value of the next-generation matrix, $ FV^{-1}$, is the quarantine reproduction number (QRN) defined as 
	\begin{equation}
		\begin{array}{l}
		\tilde{\mathcal{R}^q_k}=\rho (FV^{-1})=\dfrac{\alpha_k\beta_k\mathbf{q}_k + \beta_k\rho_k\sigma_k+\alpha_k\beta_k\tau_k+\mathbf{r}_k\beta_k\rho_k}{\mathbf{p}_k(\dot{\mathbf{q}}_k\mathbf{r}_k+\tilde{\mathbf{r}}_k\sigma_k)}
		\end{array}
		\label{eqn:Quarantine}
	\end{equation}
 where 
 $$\tilde{\mathbf{r}}_k=\phi_k+\delta_k+\mu,~ \dot{\mathbf{q}}_k=\gamma_k +\delta_k+\mu, \text{ and $k=c$ or $i$}.$$
The QRN, $\tilde{\mathcal{R}^q_k}$, gives the secondary infections generated by an individual infected with a single pathogen in a population when the fractions of exposed and infected individuals are restricted. Applying the long-established Theorem 2 in \cite{van2002reproduction}, we summarize these results in the following lemma:
\begin{lemma}
The DFE of the model \cref{eqn:MainModel_Reduced} always exists. If  $\tilde{\mathcal{R}^q_k}<1$ hold, the $\mathbf{E}^0_k$ is locally asymptotically stable ($LAS$) or unstable otherwise.
			\label{lem:DFE_Reduced}
	\end{lemma}

Results stated in \cref{lem:DFE_Reduced} imply the pathogen can be removed from the population if $\tilde{\mathcal{R}^q_k}<1$. When the threshold exceeds unity, the pathogen can invade the disease-free state. That is, the  $\mathbf{E}^0_k$ will change its stability from stable to unstable, and the new positive endemic will appear if the initial size of the infectious is sufficiently large. Ensuring the removal of the pathogen in the community does not depend on the initial conditions showing that the DFE is globally asymptotically stable is necessary \cite{zhou2002global}. 

\subsubsection{Global Stability of the DFE-(\texorpdfstring{$\mathbf{E}^{0}_{k}$)}{Lg}}

We analyze the global stability of the disease-free-equilibrium, $\mathbf{E}^{0}_{k}$, using the approach as stated by \cite{castillo2002computation}.  In the form of Equation 3.1 in \cite{castillo2002computation}, we write system \cref{eqn:MainModel_Reduced} as: 
\begin{equation}
\begin{aligned}
	\dfrac{d\mathbf{x}}{dt}=F(\mathbf{x}, \mathbf{I}), 
	\dfrac{d\mathbf{I}}{dt}=G(\mathbf{x}, \mathbf{I}), 
	\label{eqn:SSM_Global}
\end{aligned}
\end{equation}

where  $\mathbb{R}^2\ni \mathbf{x}=(S, R_k)^T$ (with $T$ denoting transpose), is a vector of uninfected individuals, and $\mathbb{R}^3\ni \mathbf{I}=(E_k, I_k, Q_k)^T$, is a vector of infected individuals. We denote  the DFE , $\mathbf{E}^0_k=(\mathbf{x}^*,0,0,0)$  where $\mathbf{x}^*$ define the DFE of system ${d\mathbf{x}}/{dt}$. Moreover, we state the following conditions:
\begin{itemize}
	\item [\textbf{H1:}] For $F(\mathbf{x}, \mathbf{I})|_{\mathbf{x}^*}$, $\mathbf{x}^*$ is globally asymptotically stable ($GAS$),
	\item [\textbf{H2:}] $G(\mathbf{x}, \mathbf{I})=A\mathbf{I}-\hat{G}(\mathbf{x}, \mathbf{I})$, $\hat{G}(\mathbf{x}, \mathbf{I})\geq 0 $ for $(\mathbf{x}, \mathbf{I})\in\mathcal{C}_k$,
\end{itemize}

where $A=G(\mathbf{x}^*, 0)$ is a Metzler matrix. If the system  \cref{eqn:MainModel_Reduced} satisfies the two assumptions, the following assertion result holds.

\begin{theorem}
	The DFE  $\mathbf{E}^0_k=({\mathbf{x}^*},0,0,0)$ of system \cref{eqn:SSM_Global}, equivalent to \cref{eqn:MainModel_Reduced} is $GAS$ if assumptions \textbf{H1} and \textbf{H2} holds and that $\tilde{\mathcal{R}^q_k} < 1$ ($LAS$).
\end{theorem}
\begin{proof} We have shown in \cref{ssst:SSM_DFE} that $\tilde{\mathcal{R}^q_k} < 1$ ($LAS$), thus, we now prove for assumptions \textbf{H1} and \textbf{H2} only.  From \cref{eqn:SSM_Global}, we have 
\begin{equation*}
\begin{array}{l}
F(\mathbf{x}, 0)= \begin{pmatrix}
\Lambda -\mu S\\
0 
\end{pmatrix}, ~~ A= \begin{pmatrix}
-\mathbf{p}_k &\beta_k &\beta_k\\ 
\rho_k & -\mathbf{q}_k& \tau_k\\
\alpha_{k} &\sigma_{k}& -\mathbf{r}_k\\
\end{pmatrix},   \\
 \text{ and, } \\
  \hat{G}(\mathbf{x}, \mathbf{I})=\begin{pmatrix}
\beta_kI_k(1-\frac{S}{N})+\beta_kQ_k(1-\frac{\Omega_{k}S}{N})\\
0\\
0
\end{pmatrix}.
\end{array}
\end{equation*}

Since $\Omega_k\in(0,1]$ and $0\leq \Omega_k S\leq S\leq N$ then $ 	\hat{G}(\mathbf{x}, \mathbf{I})\geq 0$ (\textbf{H2} holds).  Moreover,
\begin{align*}
\lim\limits_{t\to\infty}F(\mathbf{x}(t), \mathbf{I}(t))|_{\mathbf{x}^*}=\lim\limits_{t\to\infty}F(\mathbf{x}(t), 0)=\Big(\frac{\Lambda}{\mu}, 0\Big)=\mathbf{x}^*, \text{\textbf{ H1} holds} .
\end{align*}
\end{proof}
We state these results in the subsequent proposition:
\begin{proposition}
	System \cref{eqn:MainModel_Reduced} has the DFE,  $\mathbf{E}^0_k=(\Lambda/\mu,0,0,0,0)$. Whenever $\tilde{\mathcal{R}^q_k}\leq 1$ the $\mathbf{E}^0_k$  is $GAS$. Otherwise, a unique positive endemic equilibrium is stable if $\tilde{\mathcal{R}^q_k}>1$.
\end{proposition}
		
\subsubsection{Endemic equilibrilium of the SSM}	\label{ssst:SSM endemic equilibrilium}
	We now settle the strain-$c$ (and $i$) endemic equilibria of the model \cref{eqn:MainModel_Reduced}. We denoted this equilibrium by 
\begin{align*}
	\tilde{\mathbf{E}}_k^{*} =\left[\dfrac{S_k^{*}}{{N}_k^{*}},\dfrac{E_k^{*}}{{N}_k^{*}},\dfrac{I_k^{*}}{{N}_k^{*}},\dfrac{Q_k^{*}}{{N}_k^{*}}, \dfrac{R_k^{*}}{{N}_k^{*}}\right]=\left[ \tilde{S_k^*}, \tilde{E_k^*}, \tilde{I_k^*}, \tilde{Q_k^*}, \tilde{R_k^*}\right].
	\label{eqn:InfCovid_Strain_Imperfect}
\end{align*}
We express the variables $ \tilde{S_k^*}, \tilde{E_k^*}, \tilde{I_k^*}, \tilde{Q_k^*}$ and $\tilde{R_k^*}$  in terms of  $	\tilde{ \xi_i^*}$ as

	\begin{equation}
		\begin{array}{lll}
			\tilde{S_k^*}=\dfrac{\Lambda} {\tilde{\xi_{k}^*}+\mu},& 
			\tilde{E_k^*}=\dfrac{\Lambda}{\mathbf{p}_k}\dfrac{\tilde{\xi_{k}^*}} {\tilde{\xi_{k}^*}+\mu},&
			\tilde{I_k^*}=W_{1k} \dfrac{\tilde{\xi_{k}^*}} {\tilde{\xi_{k}^*}+\mu},\\
			\tilde{Q_k^*}=W_{2k} \dfrac{\tilde{\xi_{k}^*}} {\tilde{\xi_{k}^*}+\mu},&
			\tilde{R_k^*}=\dfrac{\phi_k}{\mu}	\tilde{Q _k^*}+\dfrac{\gamma_k}{\mu}	\tilde{I_k^*},&
			{N_k^*}= \dfrac{\Lambda}{\mu}\dfrac{\xi_{k}^*} {\xi_{k}^*+\mu}(1-\tilde{\Phi}_k)
			\label{eqn:InflCovidEquilibria_WithForce}
		\end{array}
	\end{equation}
	where;
	\begin{align*}
		\begin{array}{ll}
		W_{1k}=\left[\dfrac{\mathbf{r}_k\rho_k\Lambda+\alpha_k\tau_k\Lambda}{\mathbf{p}_k(\dot{\mathbf{q}}_k\mathbf{r}_k+\tilde{\mathbf{r}}_k\sigma_k)} \right],&
		W_{2k}=\left[\dfrac{\mathbf{q}_k\alpha_k\Lambda+\sigma_k\rho_k\Lambda}{\mathbf{p}_k(\dot{\mathbf{q}}_k\mathbf{r}_k+\tilde{\mathbf{r}}_k\sigma_k)} \right],\\\\
			\tilde{\Phi}_k=\left[\dfrac{\alpha_k\delta_k\mathbf{q}_k+\delta_k\mathbf{r}_k\rho_k + \delta_k\rho_k\sigma_k+\alpha_k\delta_k\tau_k}{\mathbf{p}_k(\dot{\mathbf{q}}_k\mathbf{r}_k+\tilde{\mathbf{r}}_k\sigma_k)}\right], &\tilde{ \xi_k^*}=\dfrac{\beta_k({I}_k^*+(1-\omega_{k}{I}_k^*){Q}_k^*)}{N_k^*}.\\
		\end{array}
	\end{align*}
	Let us now eliminate $	\tilde{I_k^*}$,  $	\tilde{Q_k^*}$ and $N_k^*$ from the expression of $\tilde{ \xi_k^*}$ with their corresponding expressions in \cref{eqn:InflCovidEquilibria_WithForce} to have the following cubic  equation:
 \begin{align}
	\tilde{\xi_{k}^*}(a_2\tilde{\xi_{k}^*}^2+a_1\tilde{\xi_{k}^*} +a_0) = 0,
		\label{eqn:SingleStrain_Imperfect}
	\end{align}
 where;
	\begin{align*}
		a_2 & =  (1-\tilde{\Phi}_k)\Lambda,\\
		a_1 & = W_{1k}W_{2k}\omega_{k}\beta_k\mu+ (1-\tilde{\Phi}_k)\Lambda\mu +(1-\tilde{\mathcal{R}}_q^k)\Lambda\mu,\\
		a_0 & = \mu^2(1-\tilde{\mathcal{R}}_q^k)\Lambda.
	\end{align*}
\noindent By inspection, it is easy to see that the one root is $\tilde{\xi_{k1}^*}=0$,  a disease-free equilibrium. To get the other two roots, we solve the quadratic equation inside the brackets of Equation \cref{eqn:SingleStrain_Imperfect}.  

\begin{figure}[ht]
	\centering
	\includegraphics[width=0.7\linewidth]{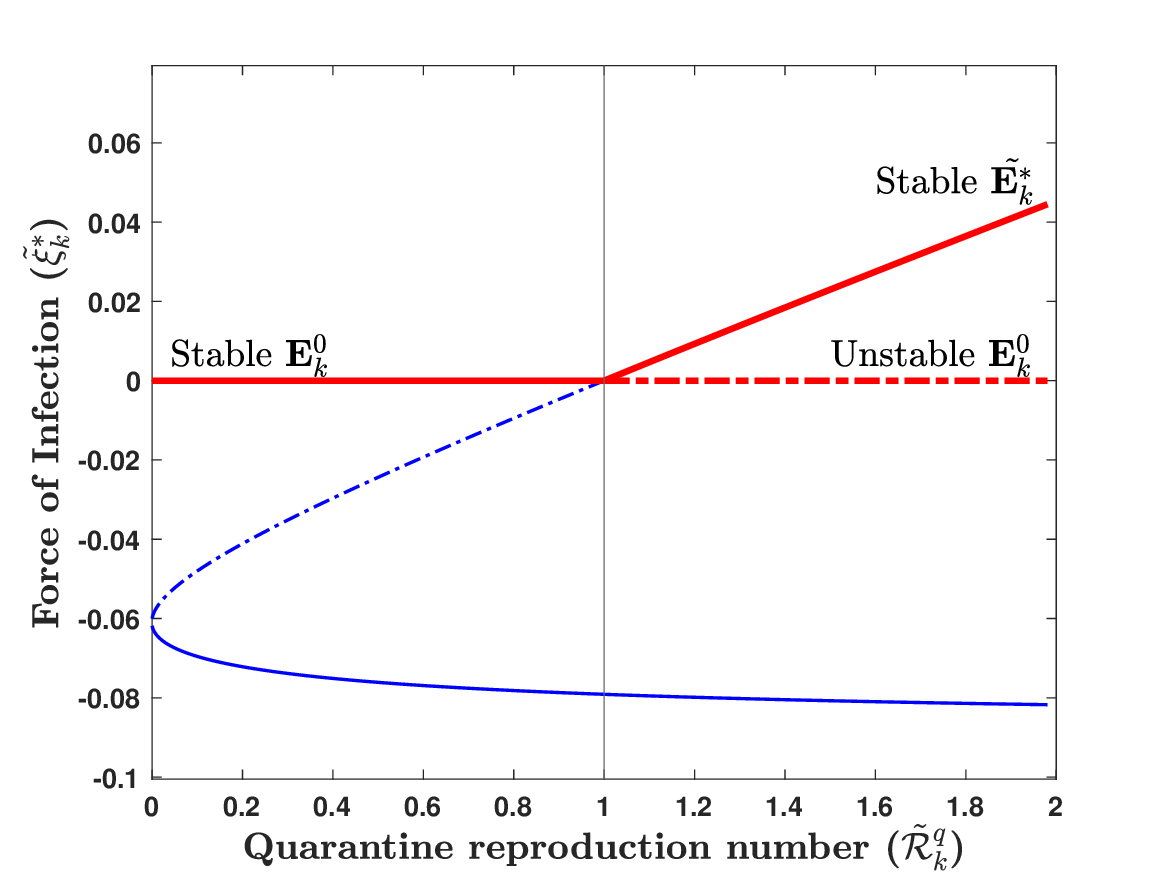}
	\caption{Schematic graphical solution $\tilde{\xi_{k}^*}$ of \cref{eqn:SingleStrain_Imperfect} with respect to the bifurcation parameter $\tilde{\mathcal{R}_k^q}$. Thick and dotted lines denote the stable and unstable equilibria, respectively. For $\tilde{\mathcal{R}_k^q}<1$, there is no positive equilibrium and the disease-free $\mathbf{E}_k^0$ is stable. For $\tilde{\mathcal{R}_k^q}>1$, there are two equilibria, of which one is positive and the disease-free $\mathbf{E}_k^0$ is unstable. Parameters used are $\beta_k\in[0.0001,6]$, $\omega_{k} = 0.6$, $\Lambda = 7.2$, $\alpha_k= 1$, $\sigma_k = 2$, $\tau_k=0.09$, $\phi_k = 5$, $\mu = 0.06$, $\delta_k = 0.09$, and $\gamma_k= 0.1$.}
	\label{fig:Bif_4CofInf_Imperfect}
\end{figure}

Using the Routh–Hurwitz conditions on the second order polynomial \cite{murray2002mathematical}, we analyze the quadratic \cref{eqn:SingleStrain_Imperfect} for possible steady states solutions around $\tilde{\mathcal{R}_k^q}$. \textit{It follows that whenever $a_i<0$, $i=0,1,2$, all roots of the \cref{eqn:SingleStrain_Imperfect} are non-negative or have positive real-parts}. The parameter  $\tilde{\Phi}_k$ gives the total proportion of people who will die from the disease. From a practical point of view, not all individuals in the population will die from the disease, so it is reasonable to assume that $\tilde{\Phi}_k\in(0,1)$: implies $a_2$ is strictly positive. If  $\tilde{\mathcal{R}_k^q}=1$, in this case $a_0=0$ and $a_1>0$. This implies the quadratic of  \cref{eqn:SingleStrain_Imperfect} has a single negative root, namely, $\tilde{\xi_{k}^*}=-a_1/a_2$. Moreover, if $\tilde{\mathcal{R}_k^q}>1$,  mathematically in this case,  the quadratic has two real roots, $(-a_1\pm \sqrt{a_1^2+4a_2a_0})/2a_2$: one root is positive, of course, this makes sense biologically. It gives a stable disease state. 
The second root is negative. Mathematically, it is defined, but, biologically is paradoxical. Finally, $a_0$ and $a_1$ are positive if $\tilde{\mathcal{R}_k^q}$ is less than unity This imply that no positive root(s) exists whenever $\tilde{\mathcal{R}_k^q}<1$. Consequently, no backward bifurcation. Plotting the ($\tilde{\mathcal{R}_k^q}, \tilde{\xi_{k}^*}$) relationship (see \cref{fig:Bif_4CofInf_Imperfect}) provides a means of determining possible steady-states and the forward bifurcation at $\tilde{\mathcal{R}_k^q}=1$. We summarize the results in the following theorem.

	\begin{theorem} Model \cref{eqn:MainModel_Reduced}, there exists a unique and positive endemic equilibrium (given by \cref{eqn:InflCovidEquilibria_WithForce}) if and only if $\tilde{\mathcal{R}_k^q}>1$. Otherwise the DFE is $GAS$. \label{thrm:InflCovi Endemic Equilibria}
	\end{theorem}
		
\subsubsection{Stability and Bifurcation  of the SSM with imperfect quarantine }
\label{ssst:Single_Strain_imperfect _quarantine}
For a rigorous proof of local stability and non-existence of backward bifurcation of the endemic equilibrium established in  \cref{thrm:InflCovi Endemic Equilibria}, we implement the Centre Manifold Theory as described by Castillo-Chavez and Song \cite{castillo2004dynamical}. For notations simplicity  we set $S= x_1$, $E_k= x_2$, $I_k= x_3$, $Q_k= x_4$, and $R_k= x_5$, so that $N_k=\sum_{n=1}^{5}x_n$. In vector form with $X=(x_1,x_2,x_3,x_4,x_5)^T$ and $\frac{dX}{dt}=G=(g_1,g_2,g_3,g_4,g_5)^T$, we can write system \cref{eqn:MainModel_Reduced} as\begin{align}
	G(\Lambda,\mu,\rho_k,\alpha_k,\sigma_k, \tau_k,\phi_k,\gamma_k, \mathbf{p}_k,\mathbf{q}_k,\mathbf{r}_k)=
	\begin{pmatrix}
		\Lambda -\xi_k x_1 - \mu x_1\\
		\xi_k x_1 -\mathbf{p}_kx_2 \\
		\rho_k x_2+\tau_k x_4-\mathbf{q}_kx_3\\
		\alpha_k x_2-\mathbf{r}_kx_4+\sigma_k x_3\\
		\phi_k x_4+\gamma_k x_3-\mu x_5
	\end{pmatrix}.
	\label{eqn:MainModel_Reduced_Transformed}
\end{align}
Jacobian matrix of system \cref{eqn:MainModel_Reduced_Transformed} about the steady-state $\mathbf{E}^0_k = (\frac{\Lambda}{\mu},0,0,0,0)$  is given by

\begin{align}
	J(\mathbf{E}^0)|_{\beta_k^*}=
	\begin{pmatrix}
		-\mu& 0&-\beta_k^*&-\beta_k^*&0\\
		0& -\mathbf{p}_k&\beta_k^*&\beta_k^*&0\\
		0& \rho_k&-\mathbf{q}_k&\tau_k&0\\
		0& \alpha_k&-\sigma_k&-\mathbf{r}_k&0\\
		0& 0&-\gamma_k&\phi_k&-\mu\\
	\end{pmatrix}.
	\label{matr:Jacobian_Transformed}
\end{align}
where $\beta_k^*$ is the bifurcation parameter defined as,
\begin{align*}\beta_k^*=
	\dfrac{\mathbf{p}_k(\dot{\mathbf{q}}_k\mathbf{r}_k+\tilde{\mathbf{r}}_k\sigma_k)}{\alpha_k\mathbf{q}_k + \rho_k\sigma_k+\alpha_k\tau_k+\mathbf{r}_k\rho_k}.
\end{align*}
With $\beta^*_k$, we revealed that the Jacobian matrix \cref{matr:Jacobian_Transformed} has a simple zero eigenvalue and that all other eigenvalues are negative. Further, a little algebra shows that the right eigenvectors of  \cref{matr:Jacobian_Transformed}  associated with zero eigenvalues given by $w=[w_1,w_2,w_3,w_4,w_5]^T$ are;
\begin{align*}
	w_1=\dfrac{-b_1}{b_5}, ~~~~~w_2=\dfrac{\mu b_1}{\mathbf{p}_kb_5},~~~~~ w_3=\dfrac{\mu b_3}{b_5}, ~~~~~w_4=\dfrac{\mu b_2}{b_5},~~~~~ w_5=1.
\end{align*}
Similarly, the left eigenvector associated with a zero eigenvalue at $\beta_k^*$  given by $$v=[v_1,v_2,v_3,v_4,v_5]$$ are;

\begin{align*}
	v_1=0,~~~~~ v_2 =\dfrac{(b_2 + b_3)v_3}{b_4},~~~~~ v_3=v_3>0,~~~~~ v_4=\dfrac{\mathbf{q}_k+\tau_k}{\sigma_k+\mathbf{r}_k}, ~~~~~v_5=0,
\end{align*}
where,
\begin{align*}
	&b_1={\mathbf{p}_k(\dot{\mathbf{q}}_k\mathbf{r}_k+\tilde{\mathbf{r}}_k\sigma_k)},~~~~~
	b_2 = \alpha_k\mathbf{q}_k  + \rho_k\sigma_k,~~~~~~~~
	b_3=\alpha_k\tau_k+\mathbf{r}_k\rho_k,~~~~~~~~~\\
	&	b_4 = \sigma_k\mathbf{p}_k+\mathbf{r}_k\mathbf{p}_k.~~~~~~~~~~~ b_5=\phi_kb_2+\gamma_kb_3.
\end{align*}

\subsubsection{Computation of bifurcation coefficients}
Despite the non-zero values vectors $v_3$ and $v_4$, the second derivatives of their corresponding functions are zero. Thus, $g_2$ is the only function required to determine the sign 
of $a$ and $b$. The non-zero second partial derivatives of $g_2$ evaluated at the DFE are: 
\begin{equation}
	\begin{aligned}
		\begin{array}{lll}
			\dfrac{\partial^2g_2}{\partial x_2\partial x_3}=- \dfrac{\mu\beta_k^*}{\Lambda}, &
			\dfrac{\partial^2g_2}{\partial x_3^2}= - \dfrac{2\mu\beta_k^*}{\Lambda}, & \\
			\dfrac{\partial^2g_2}{\partial x_3\partial x_4}= -\beta_k^*\omega_k- \dfrac{2\mu\beta_k^*}{\Lambda}, &
			\dfrac{\partial^2g_2}{\partial x_3\partial x_5}= - \dfrac{\mu\beta_k^*}{\Lambda},&
			\dfrac{\partial^2g_2}{\partial x_3\partial \beta_k^*}= 1.
		\end{array}
	\end{aligned}
	\label{eqn:ForBifurcationSign}
\end{equation}
Hence, the coefficients 
\begin{equation*}
	\begin{aligned}
		a &= \sum_{l,i,j=1}^{5}v_lw_iw_j\dfrac{\partial^2g_l}{\partial x_i\partial x_j}(0,0),\\
		&=-\dfrac{(b_2 + b_3)}{b_4}v_3w_3\mu\beta_k^*\Bigg(\dfrac{ b_1}{\Lambda\mathbf{p}_kb_5}+ \dfrac{2\mu b_3}{\Lambda b_5}+ \dfrac{\mu b_2\omega_k}{b_5}+ \dfrac{2\mu b_2}{\Lambda b_5}+ \dfrac{1}{\Lambda}\Bigg)<0,
	\end{aligned}
\end{equation*}
and
\begin{equation*}
	\begin{aligned}
		b =\sum_{l,i=1}^{5}v_lw_i\dfrac{\partial^2g_l}{\partial x_i\partial \beta_k^*}(0,0)=v_2w_3\dfrac{\partial^2g_2}{\partial x_3\partial \beta_k^*}=\dfrac{(b_2 + b_3)}{b_4}v_3w_3>0.
	\end{aligned}
\end{equation*}
Since $a<0$ and $b>0$, there is no backward bifurcation. As pointed out by Castillo et. al,  \cite{castillo2004dynamical} in item (iv) of Theorem 4.1, we establish the following theorem to rephrase the outcomes on the stability of endemic equilibrium and bifurcation of the model \cref{eqn:MainModel_Reduced}.
 
\begin{theorem} Consider model \cref{eqn:MainModel_Reduced},  equivalently to \cref{eqn:MainModel_Reduced_Transformed}, and assume $\Omega_{k}\in (0,1] $.
	\begin{enumerate}
		\item  [i.]	 When the parameter $\beta_k$ shifts from negative to positive, $\mathbf{E}_k^0$ will change its stability from stable to unstable. Alike, $\tilde{\mathbf{E}}_k^*$ changes from a negative to a positive and $LAS$.

    \item  [ii.] Provided $a<0$ and $b>0$, the  model in fact has forward bifurcation at $\tilde{\mathcal{R}}_k^q = 1$. 
	\end{enumerate}
	\label{thrm:CoviBackwardBif}
\end{theorem}

\subsection{Analysis of Multiple-Strain model (MSM) with no co-infection}\label{ssst:MSM_NO_COINF}
Here, we present the mathematical investigation of the model \cref{eqn:MainModel} when $I_{ic}=0$ but $I_i\neq 0$ and $I_c\neq 0$. The consequential  model system \cref{eqn:MainModel_Reduced2}, has four equations less than the original, \cref{eqn:MainModel}:

\begin{equation}
	\begin{array}{ll}
		S'(t)=\Lambda - \xi_{i}S-\xi_{c}S - \mu S, & Q'_i(t)=\alpha_iE_i-\mathbf{r}_iQ_i+\sigma_iI_i,\\
		E'_{i}(t)= \xi_{i}S - \mathbf{p}_iE_i +\eta_cR_c\xi_i,&Q'_c(t)=\alpha_cE_c-\mathbf{r}_cQ_c+\sigma_cI_c,\\	
		E'_{c}(t)=\xi_{c}S - \mathbf{p}_cE_c+\eta_iR_i\xi_c,&R'_{i}(t)=\phi_iQ_i+\gamma_iI_i-\eta_iR_i\xi_c-\mu R_i,\\
		I'_i(t)=\rho_iE_i+\tau_iQ_i-\mathbf{q}_iI_i,&R'_{c}(t)=\phi_cQ_c+\gamma_cI_c-\eta_cR_c\xi_i-\mu R_c,\\
		I'_c(t)=\rho_cE_c+\tau_cQ_c-\mathbf{q}_cI_c,&\\
	\end{array}
	\label{eqn:MainModel_Reduced2}
\end{equation}

\subsubsection{DFE}\label{ssst:DFE_MSM}
	The disease-free equilibrilium of model \cref{eqn:MainModel_Reduced2} is given by
\begin{equation*}
	\begin{aligned}
		\ddot{\mathbf{E}}^0  & = (\ddot{S}^0, \ddot{E}_i^0, \ddot{E}_c^0, \ddot{I}_i^0, \ddot{I}_c^0, \ddot{Q}_i^0, \ddot{Q}_c^0,\ddot{R}_i^0, \ddot{R}_c^0) = \left(\dfrac{\Lambda}{\mu},0,0,0,0,0,0,0,0\right).
		\label{eqn:DFE_Reduced2}
	\end{aligned}
\end{equation*}

We calculate the reproduction number(s) and state the local stability of the disease-free equilibrium, $ \ddot{\mathbf{E}}^0$. With a similar approach as in \cref{ssst:SSM_DFE}, we found two reproduction numbers that do not depend on each other. Symbolically,  we show them as
 
 \begin{equation}
 	\tilde{\mathcal{R}^q} =\max\{\tilde{\mathcal{R}^q_c}, \tilde{\mathcal{R}^q_i}\},
 	\label{eqn:GenRepNum}
 \end{equation}
where
\begin{equation*}
\begin{array}{lll}
	\tilde{\mathcal{R}^q_c}= \dfrac{\alpha_{c}\beta_{c}\mathbf{q}_{c} + \beta_{c}\rho_{c}\sigma_{c} +\alpha_{c}\beta_{c}\tau_{c} + \mathbf{r}_{c}\beta_{c}\rho_{c}}{\mathbf{p}_c(\dot{\mathbf{q}}_c\mathbf{r}_c+\tilde{\mathbf{r}}_c\sigma_c)},& \text{ and, } & 	\tilde{\mathcal{R}^q_i}= \dfrac{\alpha_{i}\beta_{i}\mathbf{q}_{i} + \beta_{i}\rho_{i}\sigma_{i}+\alpha_{i}\beta_{i}\tau_{i}+\mathbf{r}_{i}\beta_{i}\rho_{i}}{\mathbf{p}_i(\dot{\mathbf{q}}_i\mathbf{r}_i+\tilde{\mathbf{r}}_i\sigma_i)}.
\end{array}
	\label{eqn:RepNums}
\end{equation*}
We then state the following  proposition:
\begin{proposition}
The DFE of the model \cref{eqn:MainModel_Reduced2} always exists.  If  $\max\{\tilde{\mathcal{R}^q_c}, \tilde{\mathcal{R}^q_i}\}<1$ hold, $ \ddot{\mathbf{E}}^0$ is locally asymptotically stable (LAS). Otherwise, it is unstable.
\label{prop:DFE_MainModel_Reduced}
\end{proposition}

\Cref{prop:DFE_MainModel_Reduced} suggests that if $\tilde{\mathcal{R}^q_c}<1$ and $\tilde{\mathcal{R}^q_i}>1$ then strain-$i$ will dominate  and  drives strain-$c$ to extinction. On contrarily,  if $\tilde{\mathcal{R}^q_i}<1$ and $\tilde{\mathcal{R}^q_c}>1$ then strain-$c$ will dominate  and  drives strain-$i$ to extinction. Moreover, if the reproduction number of each strain exceeds the unity ($\tilde{\mathcal{R}^q_i}>1$ and $\tilde{\mathcal{R}^q_c}>1$), then competition; strain-$c$ and strain-$i$ will compete for susceptible individuals. In this case, the strain with a large reproduction number outcompete the other. The competition will exhibit the so-called exclusive principle if the competed strain extinct \cite{martcheva2015introduction}. For the principle to satisfy the result stated in \cref{prop:DFE_MainModel_Reduced} must hold globally. We provide the proof given in \cref{sec:GAS_Coex} and use Theorem 8.2 stated in \cite{martcheva2015introduction}, to summarize the results:

\begin{proposition}\label{prop:GAS_Coex}
The DFE, $\ddot{\mathbf{E}}^0_k$ of system \cref{eqn:MainModel_Reduced2}  is globally asymptotically stable ($GAS$) if $\max\{ \tilde{\mathcal{R}}^q_c,\tilde{\mathcal{R}}^q_i\}\leq 1$. Contrarily, when $\tilde{\mathcal{R}}^q_c>1$ and or $\tilde{\mathcal{R}}^q_i>1$ strain with the largest reproduction number dominates the other and drives them to extinction. 
\end{proposition}

\subsubsection{Dominance endemic equilibrilium}	\label{ssst:dominance endemic equilibrilium}
Let $\tilde{\mathbf{E}^d_c}$  and $	\tilde{\mathbf{E}^d_i}$ be strain-$c$ and  strain-$i$ dominance equilibria of the model \cref{eqn:MainModel_Reduced2} when  $\Omega_{k} \in(0,1]$.  Starting by deducing that  $E_{i}=I_{i}=Q_{i}=R_{i}=0$. The obtained equations that satisfy $\tilde{\mathbf{E}^d_c}$ are indeed similar to the model system \cref{eqn:MainModel_Reduced} at equilibrium on replacing $c$ by $k$. Thus analyzing the resulting model  in a similar way to that used  in \cref{ssst:SSM endemic equilibrilium}, we have 

		\begin{equation}
			\tilde{\mathbf{E}^d_c} = \left[\tilde{S_c^d},0,\tilde{E_c^d},0, 0,\tilde{I_c^d},0,0, \tilde{Q_c^d},0,0,\tilde{ R_c^d},0\right].
			\label{eqn:Covid_Endemic_Strain_Imperfect}
		\end{equation}
	
\noindent	 Similarly, in the case when $E_c=I_c = Q_c=R_c=0$ we obtain the strain-$i$ dominance equilibrium; namely, 
		\begin{align}
			\tilde{\mathbf{E}^d_i} = \left[\tilde{S_i^d},\tilde{E_i^d},0, 0,\tilde{I_i^d},0,0, \tilde{Q_i^d},0,0,\tilde{ R_i^d},0,0\right].
			\label{eqn:Influenza_Endemic_Strain_Imperfect}
		\end{align}
	
	\noindent Note, the non-zero variables:  $ \tilde{S_c^d}, \tilde{E_c^d}, \tilde{I_c^d}, \tilde{Q_c^d}$ and $\tilde{R_c^d}$ in Equation 	\cref{eqn:Covid_Endemic_Strain_Imperfect} and  $ \tilde{S_i^d}, \tilde{E_i^d}, \tilde{I_i^d}, \tilde{Q_i^d}$ and $\tilde{R_i^d}$ in Equation  \cref{eqn:Influenza_Endemic_Strain_Imperfect} are solutions obtained by replacing them with the corresponding expressions of $ \tilde{S_k^*}, \tilde{E_k^*}, \tilde{I_k^*}, \tilde{Q_k^*}$, and  $\tilde{R_k^*}$ presented in \cref{eqn:InflCovidEquilibria_WithForce} when $k=c$ or $k=i$. 

\subsubsection{Invasion reproduction numbers-(IRN)}
\label{ssst:IRN_Imperfect}
To investigate the local stability of $\tilde{\mathbf{E}}_k^d$,  the endemic equilibrium in the absence of one pathogen, we compute the invasion reproduction numbers for strain-$i$ (and strain-$c$), $_i\tilde{\mathcal{R}}_c^q$ (and $_c\tilde{\mathcal{R}}_i^q$), respectively. The quantity $_i\tilde{\mathcal{R}}_c^q$ (or $_c\tilde{\mathcal{R}}_i^q$) is used to measure the ability of strain-$i$ (or strain-$c$) to invade a system at the equilibrium of the strain-$c$ (or strain-$i$). See, for example, \cite{ martcheva2015introduction, van2002reproduction} described various approaches to calculate the IRN. The method we use to obtain $_i\tilde{\mathcal{R}}_c^q$ is the next-generation approach (included in the \cref{app_sec:IRN_Imperfect}). The expression of $_i\tilde{\mathcal{R}}_c^q$, simplifies to
	\begin{align}
		_i\tilde{\mathcal{R}}_c^q&=\Big[\tilde{S_c^*}+ \tilde{R_c^*}\eta_{c}\Big] \tilde{\mathcal{R}}_i^q.
		\label{eqn:IRN_Inf_ImperfectQ}
	\end{align}
Analogously, we define  the strain-$c$ invasion reproduction number, $_c\tilde{\mathcal{R}}_i^q,$ under the assumption that the strain-$i$ is resident to obtain
\begin{equation}
	\begin{aligned}
		_c\tilde{\mathcal{R}}_i^q&= \Big[\tilde{S_i^*}+\tilde{R_i^*}\eta_{i}\Big]\tilde{\mathcal{R}^q_c}.
	\end{aligned}
	\label{eqn:IRN_Cov_ImperfectQ}
\end{equation}
 From Equation \cref{eqn:IRN_Inf_ImperfectQ},  we notice that the term $\tilde{S_c^*}\tilde{\mathcal{R}}_i^q$ provides secondary cases of susceptible individuals that one individual infected with strain-$i$  can produce in the population. The term $\tilde{R_c^*}\eta_{c}\tilde{\mathcal{R}}_i^q$ provide the secondary cases one strain-$i$ infected individual can produce in proportions of the individual recovered from strain-$c$. The interpretation of Equation \cref{eqn:IRN_Cov_ImperfectQ} is treated similarly to  \cref{eqn:IRN_Inf_ImperfectQ}. We then compile the results on the existence and stability of $\tilde{\mathbf{E}^d_c}$ (and  $\tilde{\mathbf{E}^d_i}$) of the model \cref{eqn:MainModel_Reduced2} in the following proposition: 

\begin{proposition} 
	Assume that $j\neq k=c$ or $i$.	If 	$\tilde{\mathcal{R}}_k^q>1$, then the endemic equilibrium $\tilde{\mathbf{E}^d_k}$ exists. Whenever $	_j\tilde{\mathcal{R}}_k^q>1$, the equilibrium $\tilde{\mathbf{E}^d_k}$ is unstable. Otherwise, the equilibrium $\tilde{\mathbf{E}^d_k}$ is $LAS$ if the invasion reproduction number  $_j\tilde{\mathcal{R}}_k^q<1$.
\end{proposition}

 \subsubsection{Coexistence of endemic  with no co-infection}\label{sec:Coexistence_Imp}
To obtain the analytical solution, we assume complete protection of recovered individuals against the secondary pathogen,  that is,  $\eta_c=\eta_i=0$.   
\noindent Let $\ddot{\mathbf{E}}^*$ be the coexistence equilibrium. In terms of the force of infections, $\ddot{\xi_i^*} $ and $\ddot{\xi_c^*} $, defined as
\begin{equation}
\begin{array}{lll}
	\ddot{\xi_i^* }= \dfrac{\beta_i(\ddot{I_{i}}+(1-\omega_{i}\ddot{I_{i}})\ddot{Q_{i}})}{\ddot{N_t}}, & \text{ and } & \ddot{\xi_c^*} = \dfrac{\beta_c(\ddot{I_{c}}+(1-\omega_{c}\ddot{I_{c}})\ddot{Q_{c}})}{\ddot{N_t}}
\end{array}
\label{eqn:Coexistence_4CofInf}
\end{equation}
 where $\ddot{N_t} = ({\Lambda}-(\delta_{i}(\ddot{I_i}+\ddot{Q_i}))-\delta_{c}(\ddot{I_c}+\ddot{Q_c}))/{\mu}$, the non-zero variables of the  equilibrium 
\begin{align*}
\ddot{\mathbf{E}}^* = \Bigg[\dfrac{\ddot{S}}{\ddot{N_t}},\dfrac{\ddot{E_{i}}}{\ddot{N_t}},\dfrac{\ddot{E_{c}}}{\ddot{N_t}},0,\dfrac{\ddot{I_{i}}}{\ddot{N_t}}\dfrac{\ddot{I_{c}}}{\ddot{N_t}},0,\dfrac{\ddot{Q_{i}}}{\ddot{N_t}},\dfrac{\ddot{Q_{c}}}{\ddot{N_t}},0,\dfrac{\ddot{R_{i}}}{\ddot{N_t}}, \dfrac{\ddot{R_{c}}}{\ddot{N_t}},0\Bigg]\\=\Big[{\ddot{S}^*},\ddot{E_{i}^*},\ddot{E_{c}^*},0,\ddot{I_{i}^*},\ddot{I_{c}^*},0,\ddot{Q_{i}^*},\ddot{Q_{c}^*},0,\ddot{R_{i}^*}, \ddot{R_{c}^*},0\Big],
\end{align*}
satisfying the successive equations:

\begin{equation}
\begin{array}{lll}
\ddot{S^*}=\dfrac{\Lambda}{	\ddot{\xi_i^*}+	\ddot{\xi_c^*}+\mu},&\ddot{I_{i}^*}=\dfrac{\ddot{\xi_i^*}W_{11}}{(\ddot{\xi_i^*}+	\ddot{\xi_c^*}+\mu)},&\ddot{Q_{c}^*}=\dfrac{\ddot{\xi_c^*}W_{44}}{(\ddot{\xi_i^*}+	\ddot{\xi_c^*}+\mu)},\\
\ddot{E_{i}^*}=\dfrac{\ddot{\xi_i^*}\Lambda}{\mathbf{p}_i(\ddot{\xi_i^*}+	\ddot{\xi_c^*}+\mu)},&\ddot{I_{c}^*}=\dfrac{\ddot{\xi_c^*}W_{33}}{(\ddot{\xi_i^*}+	\ddot{\xi_c^*}+\mu)},&\ddot{R_{i}^*}=\dfrac{\phi_i}{\mu}\ddot{Q_{i}^*}+\dfrac{\gamma_i}{\mu}\ddot{Q_{i}^*},\\
\ddot{E_{c}^*}=\dfrac{\ddot{\xi_c^*}\Lambda}{\mathbf{p}_c(\ddot{\xi_i^*}+	\ddot{\xi_c^*}+\mu)},&\ddot{Q_{i}^*}=\dfrac{\ddot{\xi_i^*}W_{22}}{(\ddot{\xi_i^*}+	\ddot{\xi_c^*}+\mu)},&\ddot{R_{c}^*}=\dfrac{\phi_c}{\mu}\ddot{Q_{c}^*}+\dfrac{\gamma_c}{\mu}\ddot{Q_{c}^*}.\\
\end{array}
\label{eqn:Coexistence_EndemicV}
\end{equation}
The parameter $W_{11}=W_{1k}$ when ($k=i$), $W_{22}=W_{2k}$ when ($k=i$), $W_{33}=W_{1k}$ when ($k=c$),  and $W_{44}=W_{2k}$ when ($k=c$). 
Employing \cref{eqn:Coexistence_EndemicV} we obtain, from \cref{eqn:Coexistence_4CofInf},
\begin{equation}
	\begin{array}{ll}
		0 =& \ddot{\xi_{i}^*}\Big[W_{55}\omega_i\beta_i\mu\ddot{\xi_i^*} +(1-\Phi_c)\ddot{\xi_c^*}^2 +(1-\Phi_i)\ddot{\xi_i^*}^2 + \Big[(1-\Phi_c)+(1-\Phi_i)\Big]\ddot{\xi_i^*}\ddot{\xi_c^*}\\
		&+(1-\Phi_c)\mu\ddot{\xi_c^*}+(1-\Phi_i)\mu\ddot{\xi_i^*}+(1-\tilde{\mathcal{R}}_i^q)\mu\ddot{\xi_c^*}+(1-\tilde{\mathcal{R}}_i^q)\mu\ddot{\xi_i}+(1-\tilde{\mathcal{R}}_i^q)\mu^2 \Big]\\
		& = f_1(\ddot{\xi_{c}^*},\ddot{\xi_{i}^*}), \\
		0 =&  \ddot{\xi_{c}^*}\Big[W_{66}\omega_c\beta_c\mu\ddot{\xi_c^*}  +(1-\Phi_i)\ddot{\xi_i^*}^2 +(1-\Phi_c)\ddot{\xi_c^*}^2+ \Big[(1-\Phi_i)+(1-\Phi_c)\Big]\ddot{\xi_i^*}\ddot{\xi_c^*}\\
		&+(1-\Phi_c)\mu\ddot{\xi_c^*}+(1-\Phi_i)\mu\ddot{\xi_i^*}+(1-\tilde{\mathcal{R}}_c^q)\mu\ddot{\xi_c^*}+(1-\tilde{\mathcal{R}}_c^q)\mu\ddot{\xi_i}+(1-\tilde{\mathcal{R}}_c^q)\mu^2 \Big]\\
		& = f_2(\ddot{\xi_{c}^*},\ddot{\xi_{i}^*}).
	\end{array}
\label{eqn:Coexistence_system}
\end{equation}
where,
\begin{align*}
\begin{array}{lll}
W_{55}=\left[\dfrac{(\mathbf{r}_i\rho_{i}\Lambda+\alpha_i\tau_{i}\Lambda)(\mathbf{q}_i\alpha_i+\sigma_i\rho_i)}{{\mathbf{p}_k(\dot{\mathbf{q}}_k\mathbf{r}_k+\tilde{\mathbf{r}}_k\sigma_k)}^2} \right]& \text{ and } & W_{66}=\left[\dfrac{(\mathbf{r}_c\rho_{c}\Lambda+\alpha_c\tau_{c}\Lambda)(\mathbf{q}_c\alpha_c+\sigma_c\rho_c)}{{\mathbf{p}_c(\dot{\mathbf{q}}_c\mathbf{r}_c+\tilde{\mathbf{r}}_c\sigma_c)}^2} \right].
\end{array}
\end{align*}
The fixed points, $(\ddot{\xi_{c}^*}, \ddot{\xi_{i}^*})$  are the solutions  $s_1, s_2, s_3, s_4, s_5$ of $f_1(\ddot{\xi_{c}^*},\ddot{\xi_{i}^*})=f_2(\ddot{\xi_{c}^*},\ddot{\xi_{i}^*}) =0$. The solution $s_1$ obtained when $\ddot{\xi_{c}^*}=\ddot{\xi_{i}^*}=0$. When either $\ddot{\xi_{c}^*}=0$ or $\ddot{\xi_{i}^*}=0$, the obtained solutions are $s_1, s_2$, and $s_3$, given by

\begin{equation}
	\begin{array}{lll}
		s_1 = (0,0); & s_2 = \bigg(0,\dfrac{-b_1\pm\sqrt{b_1^2-4a_1c_1}}{2a_1}  \bigg);&s_3 =  \bigg( \dfrac{-b_2\pm\sqrt{b_2^2-4a_2c_2}}{2a_2} ,0\bigg)\\
	\end{array}
	\label{eqn:Coexistence_Sol_interms_Xi_Xc}
\end{equation}
where,
\begin{equation*}
	\begin{array}{ll}
		a_1 = (1-\Phi_i); & a_2 = (1-\Phi_c);\\
		b_1= W_{55}\omega_i\beta_i\mu + (1-\tilde{\mathcal{R}^q_i}) + (1-\Phi_i); & b_2= W_{66}\omega_c\beta_c\mu+ (1-\tilde{\mathcal{R}^q_c}) + (1-\Phi_c);\\
		c_1 =  (1-\tilde{\mathcal{R}^q_i})\mu^2;& c_2 =  (1-\tilde{\mathcal{R}^q_c})\mu^2.
	\end{array}
\end{equation*}
The first fixed point $s_1=(0, 0)$ in \cref{eqn:Coexistence_Sol_interms_Xi_Xc} corresponds to the state when everyone in the population is susceptible. The solutions $s_2$ and $s_3$ correspond to strain-$i$ and strain-$c$ dominance equilibrium, respectively. For a detailed analysis of the existence and stability of such steady states, see \cref{ssst:dominance endemic equilibrilium}. Other nonzero solutions,  $s_4$ and $s_5$ when  $\ddot{\xi_{c}^*}>0$ and $\ddot{\xi_{i}^*}>0$ are described in this subsection.

\begin{figure}[htbp]
  \centering
  \label{fig:a}\includegraphics[width=0.49\textwidth]{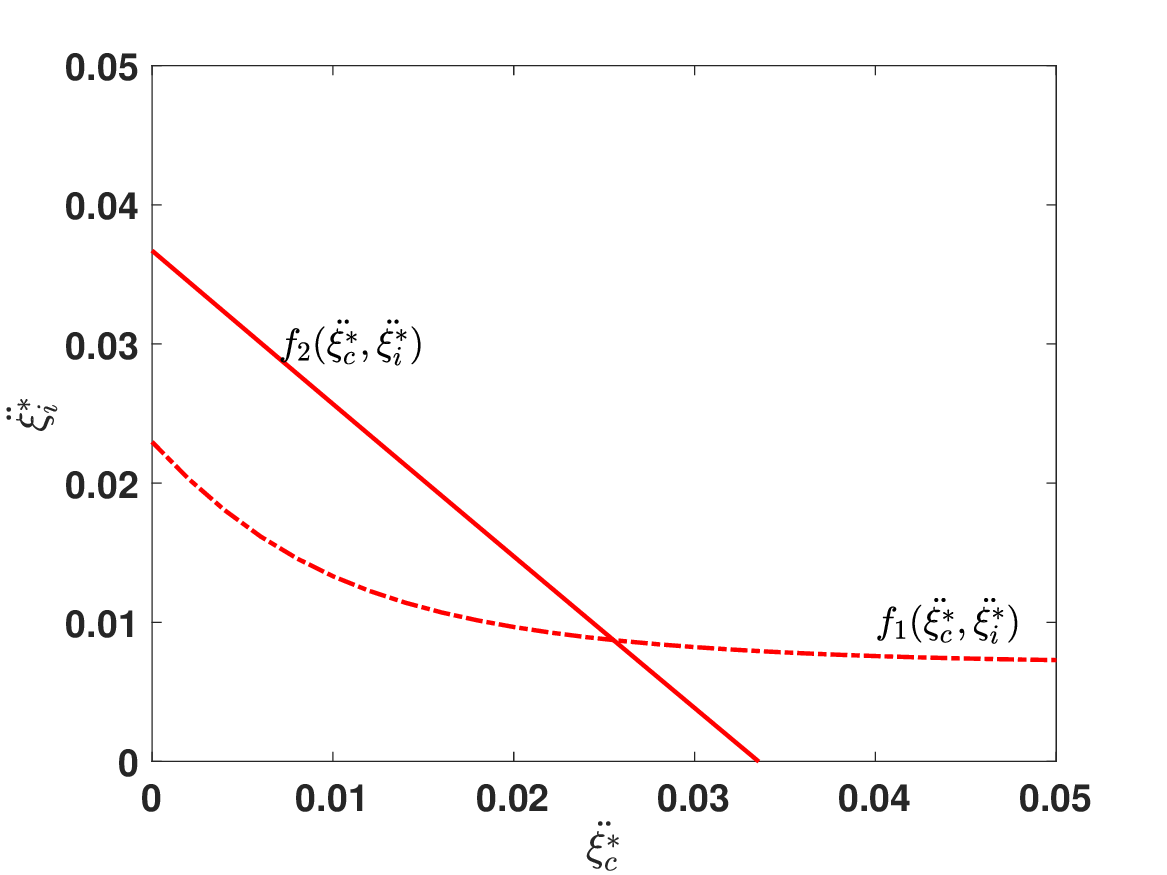}
  \label{fig:b}\includegraphics[width=0.49\textwidth]{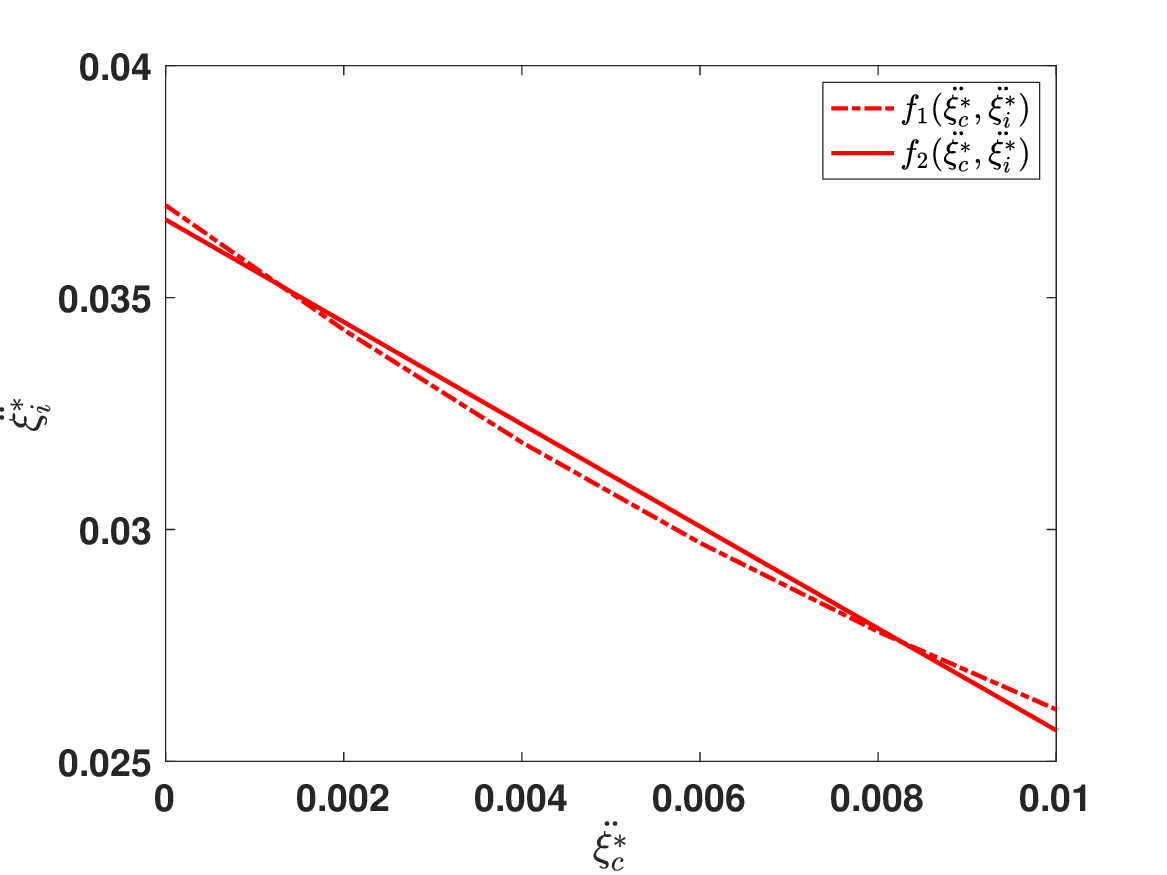}
  \caption{Possible coexistence solutions. The parameters value used for these figure are $\Lambda=0.0004$, $\beta_c=0.8$, $\rho_i=0.3$, $\rho_c=0.4$, $\alpha_i=0.8$, $\alpha_c=0.3$, $\sigma_i=.9$, $\sigma_c=.6$, $\tau_i=.1$, $\tau_c=.12$, $\delta_i=.1$, $\delta_c=.01$, $\mu=0.01$, $\delta_{ic}=.1$, $\phi_i=0.2$, $\phi_c=0.16$, $\gamma_i=0.2$, $\gamma_c=0.4$, $\omega_i=0.002$, $\omega_c=0.05$ except for the left figure $\beta _i=0.8$, and for the right figure  $\beta _i=1.1$.}
  \label{fig:coexistence}
\end{figure}

Before discussing and establish condition(s) necessary for the coexistence solution(s) in \cref{fig:coexistence}, we start by graphing solution for system \cref{eqn:Coexistence_system} in ($\ddot{\xi_{c}^*}, \ddot{\xi_{i}^*}$) relationship to check if the system \cref{eqn:Coexistence_system}  has at least one solution in the case $\ddot{\xi_{c}^*}$ and $\ddot{\xi_{i}^*}$ are both greater than zero. If there is no such solution, then there is no coexistence equilibrium. From the figure, it is easy to see that several solutions are possible depending on the parameter values.

\begin{theorem}
Suppose that each(or any) strain can invade the other population, that is $_i\tilde{\mathcal{R}}_c^q>1$, and (or) $	_c\tilde{\mathcal{R}}_i^q>1$. 
System \cref{eqn:MainModel_Reduced2} has at least one coexistence endemic equilibrium provided that the $\min\{\tilde{\mathcal{R}^q_c}, \tilde{\mathcal{R}^q_i}\}>1.$\label{thrm:Coexistence}
\end{theorem}

\begin{figure}[ht]
	\centering
	\includegraphics[width=\textwidth]{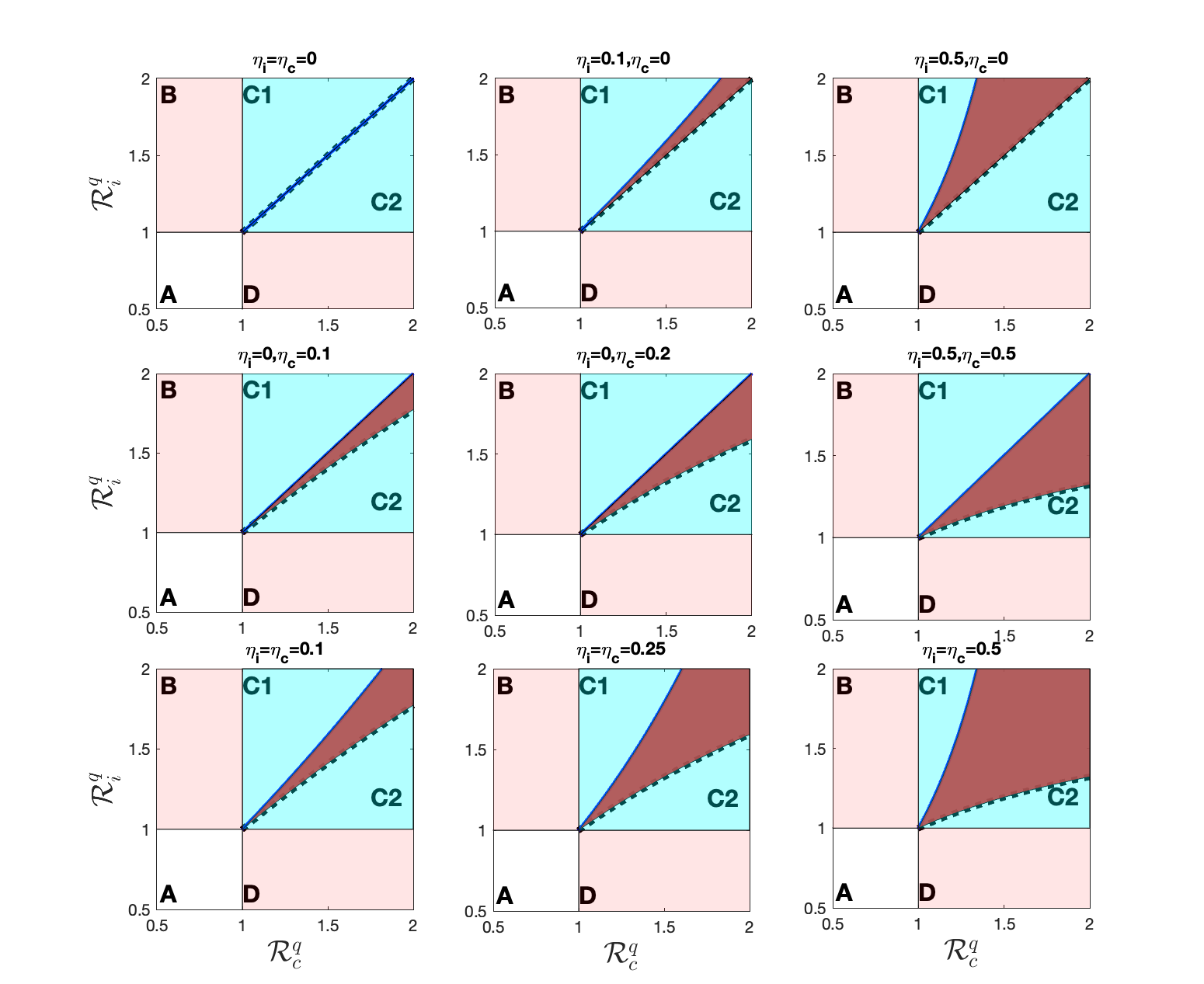}
\caption{Dominance, competition, and co-existence regions characterized quarantine reproduction number by different values of  cross-immunity, $\eta_i$ and $\eta_c$. Other parameters are taken as follows: $\Lambda=0.0004$, $\beta_c\in[0.25,1]$, $\beta _i=0.69$, $\rho_i=0.5$, $\rho_c=0.6$, $\rho_{ic}=0.5$, $\alpha_i=0.2$, $\alpha_c=0.5$, $\alpha_{ic}=0.5$, $\sigma_i=.005$, $\sigma_c=.006$, $\sigma_{ic}=.009$, $\tau_i=.05$, $\tau_c=.06$, $\tau_{ic}=.05$, $\delta_i=.001$, $\delta_c=.01$, $\mu=0.004$, $\delta_{ic}=.1$, $\phi_i=0.5$, $\phi_c=0.6$, $\gamma_i=0.2$, $\gamma_c=0.125$, $\gamma_{ic}=0.1$  }
	\label{fig:CROSS1}
\end{figure}

 \begin{proof}
Dividing each equation in \cref{eqn:Coexistence_system} by the common factor and then adding the acquired equations, we have;
\begin{align*}
	W_{55}\omega_i\beta_i\mu\ddot{\xi_i^*}-W_{66}\omega_c\beta_c\mu\ddot{\xi_c^*} +(\tilde{\mathcal{R}}_c^q-\tilde{\mathcal{R}}_i^q)\mu\ddot{\xi_c^*}+(\tilde{\mathcal{R}}_c^q-\tilde{\mathcal{R}}_i^q)\mu\ddot{\xi_i^*}+(\tilde{\mathcal{R}}_c^q-\tilde{\mathcal{R}}_i^q)\mu^2=0.
\end{align*}
We assume that $\tilde{\mathcal{R}^q_i}=\tilde{\mathcal{R}^q_c}>1$ , to obtain
\begin{equation}
	\begin{array}{lll}
		s_4 =  \bigg( \dfrac{W_{55}\omega_i\beta_i\ddot{\xi_i^*}}{W_{66}\omega_c\beta_c}, \ddot{\xi_i^*} \bigg)
		& \text{ and } & 	s_5 = \bigg( \ddot{\xi_{c}^*}, \dfrac{W_{66}\omega_c\beta_c\ddot{\xi_c^*}}{W_{55}\omega_i\beta_i} \bigg).
	\end{array}
	\label{eqn:Coexistence_Sol_intermsXiXc}
\end{equation}

\noindent It should be noted  that, $s_4$ and $s_5$ exist if and only if $\ddot{\xi_{c}^*}>0$ and $\ddot{\xi_{i}^*}>0$. Making usage of \cref{eqn:Coexistence_Sol_intermsXiXc} and \cref{eqn:Coexistence_system} while collecting like terms, we obtain after manipulations the following two quadratic equations:

\begin{equation}
\begin{aligned}
	a_{11} \ddot{\xi_{i}^*}^2 + b_{11} \ddot{\xi_{i}^*} + c_{11}&=0,\\
	a_{22} \ddot{\xi_{c}^*}^2 + b_{22} \ddot{\xi_{c}^*} + c_{22}&=0,\\
	\label{eqn:Coexistance}
\end{aligned}
\end{equation}
where
\begin{align*}
	a_{11}&=W_{66}^2\beta_{c}^2\omega_{c}^2(1-\Phi_i) + W_{55}W_{66}\beta_{c}\beta_{i }\omega_{c}\omega_{i}\Big[(1-\Phi_c) +(1-\Phi_i)\Big] + W_{55}^2\beta_{i}^2\omega_{i}^2(1-\Phi_c),\\
	b_{11}&= W_{55}W_{66}^2\beta_{c}^2\beta_{i }\omega_{c}^2\omega_{i}\mu + (1-\Phi_c)W_{55}W_{66}\beta_{c}\beta_{i }\omega_{c}\omega_{i}\mu+ (1-\Phi_i)W_{66}^2\beta_{c}^2\omega_{c}^2\mu+ \\
	&~~~~(1-\tilde{\mathcal{R}_i^q})\Big[W_{55}W_{66}\beta_{c}\beta_{i }\omega_{c}\omega_{i}\mu+ W_{66}^2\beta_{c}^2\omega_{c}^2\mu\Big],\\
	c_{11}&=W_{66}^2\beta_{c}^2\omega_{c}^2\mu^2(1-\tilde{\mathcal{R}_i^q}),
\end{align*}

besides,
\begin{align*}
a_{22}&=W_{66}^2\beta_{c}^2\omega_{c}^2(1-\Phi_i) + W_{55}W_{66}\beta_{c}\beta_{i }\omega_{c}\omega_{i}\Big[(1-\Phi_c) +(1-\Phi_i)\Big] + W_{55}^2\beta_{i}^2\omega_{i}^2(1-\Phi_c),\\
b_{22}&= W_{55}^2W_{66}\beta_{c}\beta_{i }^2\omega_{c}\omega_{i}^2\mu + (1-\Phi_i)W_{55}W_{66}\beta_{c}\beta_{i }\omega_{c}\omega_{i}\mu+ (1-\Phi_c)W_{55}^2\beta_{i}^2\omega_{i}^2\mu+ \\
&~~~~(1-\tilde{\mathcal{R}_c^q})\Big[W_{55}W_{66}\beta_{c}\beta_{i }\omega_{c}\omega_{i}\mu+ W_{55}^2\beta_{i}^2\omega_{i}^2\mu\Big],\\
c_{22}&=W_{55}^2\beta_{i}^2\omega_{i}^2\mu^2(1-\tilde{\mathcal{R}_c^q}).
\end{align*}
 \noindent We analyze the two equations in \cref{eqn:Coexistance} for possible steady states solutions around the reproduction numbers, $\tilde{\mathcal{R}_c^q}$ and $\tilde{\mathcal{R}_i^q} $. The parameters $\tilde{\Phi}_c,\tilde{\Phi}_i\in(0,1)$. Thus $a_{11}$ and $a_{22}$ are strictly positive. Applying the Routh–Hurwitz criterion, it follows that when the coefficients signs of the constant terms (i.e., $c_{11}$ and $c_{22}$) are negative, it follows that a positive root with a real part will exist. This is only possible whenever $\tilde{\mathcal{R}_c^q}>1$ and $\tilde{\mathcal{R}_i^q}>1$. 
  \end{proof}
  Note that \cref{fig:CROSS1}  is the analysis and projection of the co-exist system in \cref{eqn:MainModel_Reduced2} with cross-immunity onto the ($\tilde{\mathcal{R}^q_c}$, $\tilde{\mathcal{R}^q_i}$) plane performed to confirm theoretical results in \cref{thrm:Coexistence}. Because we assumed symmetry of the model about the two strains, we also observe that the solution is symmetrical about the line ${\mathcal{R}^q_c}={\mathcal{R}^q_i}>1$. The boundary curves (in region \textbf{C=C1+C2} ), dotted and solid lines, are given parametrically respectively by \cref{eqn:IRN_Inf_ImperfectQ} and \cref{eqn:IRN_Cov_ImperfectQ}. Moreover, the interactions between strain-$c$ and -$i$ were linked by cross-immunity parameters, $\eta_c$, and $\eta_i$  from each. By varying values of the cross-immunity, we observed that the larger $\eta_i$ and $\eta_c$ are, the larger the co-infected region. For instance,  the top left figure is simply the visualization of the coexistence region obtained by setting the cross-protection rates to zero, such that $\eta_{c}=\eta_{i}=0$. The process proceeds in two ways to obtain the remaining figures: holding one parameter to zero and increasing the other parameter or by increasing both parameters ($\eta_{c}$ and $\eta_{i}$).

Besides, from the portraiture \cref{fig:CROSS1}, we obtained various regions,  all of which have biological consequences. We call A a lose-lose region. In this region, all the reproduction numbers were less than one, meaning that, as time passes, all strains will die. In B, strain-i will win, and strain-c will lose. In D, strain-c will win, but strain-i will lose. We, therefore, call the competition outcome/behavior in B and D a Win-lose competition. Next, we looked \textbf{C}, (a Win-win region). Here the reproduction number of each strain is large than one. Intuitively this means that each pathogen expects to come out ahead. We call the situation a Win-win competition. Further, we define a draw outcome/competition. A competition if neither strain-$c$ nor strain-$i$ will lose or win the competition, regardless that their reproduction numbers exceed one, see the top left figure of \cref{fig:CROSS1}. The outcome is shown by a full-dotted line (when $\tilde{\mathcal{R}}_c^q = \tilde {\mathcal{R}}_i^q>1$).

\subsection{Existence of co-infection equilibrium}
Note that the analytical analysis of the two-strain model in \cref{ssst:MSM_NO_COINF} is marked on the assumption that $E_{ic}=I_{ic}=Q_{ic}=R_{ic}=0$.
We now consider the analysis of the model system, presented in \cref{eqn:MainModel}. \Cref{ssst:basic_properties} describe the non-negativity and boundedness of the model solution. Let us nail the stationary states. The DFE  of the model  is given by

\begin{equation}
	\begin{array}{ll}
		\mathbf{E}^0  & = (S^0, E_i^0, E_c^0, E_{ic}^0, I_i^0, I_c^0, I_{ic}^0, Q_i^0, Q_c^0, Q_{ic}^0,R_i^0, R_c^0, R_{ic}^0) \\
                &= \left(\dfrac{\Lambda}{\mu},0,0,0,0,0,0,0,0,0,0,0,0\right)\\.
		\label{eqn:DFE_MainModel}
	\end{array}
\end{equation}

\noindent Quarantine reproduction numbers are given in  Equation \cref{eqn:GenRepNum}. The local stability analysis of $\mathbf{E}^0$ is similar to that provided in \cref{ssst:DFE_MSM} except that the co-infected model \cref{eqn:MainModel} makes biological sense in $\mathcal{C}$, see \cref{ssst:basic_properties} for more details. Let \begin{align*}
\tilde{\mathbf{E}^{**}_{ic}}= \Bigg[\dfrac{\hat{S}}{\hat{N_T}},\dfrac{\hat{E_{i}}}{\hat{N_T}},\dfrac{\hat{E_{c}}}{\hat{N_T}},\dfrac{\hat{E_{ic}}}{\hat{N_T}},\dfrac{\hat{I_{i}}}{\hat{N_T}}\dfrac{\hat{I_{c}}}{\hat{N_T}},\dfrac{\hat{I_{ic}}}{\hat{N_T}},\dfrac{\hat{Q_{i}}}{\hat{N_T}},\dfrac{\hat{Q_{c}}}{\hat{N_T}},\dfrac{\hat{Q_{ic}}}{\hat{N_T}},\dfrac{\hat{R_{i}}}{\hat{N_T}}, \dfrac{\hat{R_{c}}}{\hat{N_T}},\dfrac{\hat{R_{ic}}}{\hat{N_T}}\Bigg]\\=\Big[{S^{**}},{E_i^{**}}, {E_c^{**}}, {E_{ic}^{**}},{I_i^{**}}, {I_c^{**}}, {I_{ic}^{**}},{Q_i^{**}}, {Q_c^{**}}, {Q_{ic}^{**}},{R_i^{**}}, {R_c^{**}}, {R_{ic}^{**}}\Big],
\end{align*} 
be the co-exist endemic equilibrium with individuals infected by strain-$c$ and -$i$. We have seen in  \cref{thrm:Coexistence}  that if the co-infected individuals are not present, the coexistence equilibrium, $\ddot{\mathbf{E}}^*$ occurs when $\min\{\tilde{\mathcal{R}^q_c}, \tilde{\mathcal{R}^q_i}\}>1$ and $_i\tilde{\mathcal{R}}_c^q>1$, and (or) $	_c\tilde{\mathcal{R}}_i^q>1$. A key question is: does this condition enough for the coexistence-endemic equilibrium \textit{"with"} co-infected individuals to occur? To answer this question, we evaluate the invasion reproduction numbers when $I_{ic}\neq 0$.

\subsection{Invasion reproduction number}
We use the next-generation approach to derive the strain-$i$ invasion reproduction number, $_i\tilde{\mathcal{I}}_c^q$ when strain-$c$ is resident. The infectious classes are $E_{i}, E_{ic}, I_{i} , I_{ic}, Q_{i}~\&Q_{ic}$. Procedures in \cref{app_sec:IRN_Imperfect} are repeated to get 
\begin{align}
	_i\tilde{\mathcal{I}}_c^q&=S^{**}\tilde{\mathcal{R}}_i^q + I_c^{**}\tilde{\mathfrak{R}}_i^{c} + R_c^{**}\eta_{c}\tilde{\mathcal{R}}_i^q.
	\label{eqn:IRN_Inf_ImperfectQ_Coinf}
\end{align}
Analogously, the strain-$c$ invasion reproduction number, $_c\tilde{\mathcal{I}}_i^q$ under the assumption that strain-$i$ is resident, respectively, is defined as
\begin{equation}
	\begin{aligned}
		_c\tilde{\mathcal{I}}_i^q&= S^{**}\tilde{\mathcal{R}}_c^q + I_i^{**}\tilde{\mathfrak{R}}_c^{i} + R_i^{**}\eta_{i}\tilde{\mathcal{R}}_c^q.
	\end{aligned}
	\label{eqn:IRN_Cov_ImperfectQ_Coinf}
\end{equation}

In \cref{eqn:IRN_Inf_ImperfectQ_Coinf} and \cref{eqn:IRN_Cov_ImperfectQ_Coinf}, the first and the last terms have a similar interpretation as in \cref{ssst:IRN_Imperfect}. The second terms  $\tilde{\mathfrak{R}}_i^c$ and $\tilde{\mathfrak{R}}_c^i$ gives the co-infected quarantine reproduction number (CQRN) generated by strain-$i$ and strain-$c$ in the sub-population of strain-$c$ infected (${I_c^{**}}$) and strain-$i$ infected individuals (${I_i^{**}}$), respectively. The quantities $\tilde{\mathfrak{R}}_i^{c} $ and $\tilde{\mathfrak{R}}_c^{i} $ are given by

\begin{equation*}
	\begin{aligned}
		\tilde{\mathfrak{R}}_i^{c} =\dfrac{\alpha_{ic}\beta_{i}\mathbf{q}_{ic} + \beta_{i}\rho_{ic}\sigma_{ic}+\alpha_{ic}\beta_{i}\tau_{ic}+\mathbf{r}_{ic}\beta_{i}\rho_{ic}}{\mathbf{p}_{ic}(\dot{\mathbf{q}}_{ic}\mathbf{r}_{ic}+\tilde{\mathbf{r}}_{ic}\sigma_{ic})},
	\end{aligned}
\end{equation*}
and
\begin{equation*}
	\begin{aligned}
		\tilde{\mathfrak{R}}_c^{i} =\dfrac{\alpha_{ic}\beta_{c}\mathbf{q}_{ic} + \beta_{c}\rho_{ic}\sigma_{ic} + \alpha_{ic}\beta_{c}\tau_{ic} + \mathbf{r}_{ic}\beta_{c}\rho_{ic}} {\mathbf{p}_{ic}(\dot{\mathbf{q}}_{ic}\mathbf{r}_{ic}+\tilde{\mathbf{r}}_{ic}\sigma_{ic})}.
	\end{aligned}
\end{equation*}

\noindent If $\tilde{\mathfrak{R}}_c^{i}<1$ and $\tilde{\mathfrak{R}}_i^{c}<1$ then in the long-run $I_i^{**}\tilde{\mathfrak{R}}_c^{i}\to 0$ and $I_c^{**}\tilde{\mathfrak{R}}_i^{c}\to 0$; this implies that co-infected sub-population will de-escalate.  With these results, we establish the following theorem.

\begin{theorem}
    Assume that  $\tilde{\mathcal{R}^q_i}>1$ and $\tilde{\mathcal{R}^q_c}>1$. The system \cref{eqn:MainModel} has a co-existence endemic equilibrium  provided that  $_i\tilde{\mathcal{I}}_c^q>1$ and / or $	_c\tilde{\mathcal{I}}_i^q>1$. Moreover, the equilibrium will have co-infected individuals provided that  $\tilde{\mathfrak{R}}_c^{i}>1$ and or $\tilde{\mathfrak{R}}_i^{c}>1$. 
	\label{thrm:Coinfected}
\end{theorem}

\begin{figure}[htbp]
  \centering
	\includegraphics[width=0.45\textwidth]{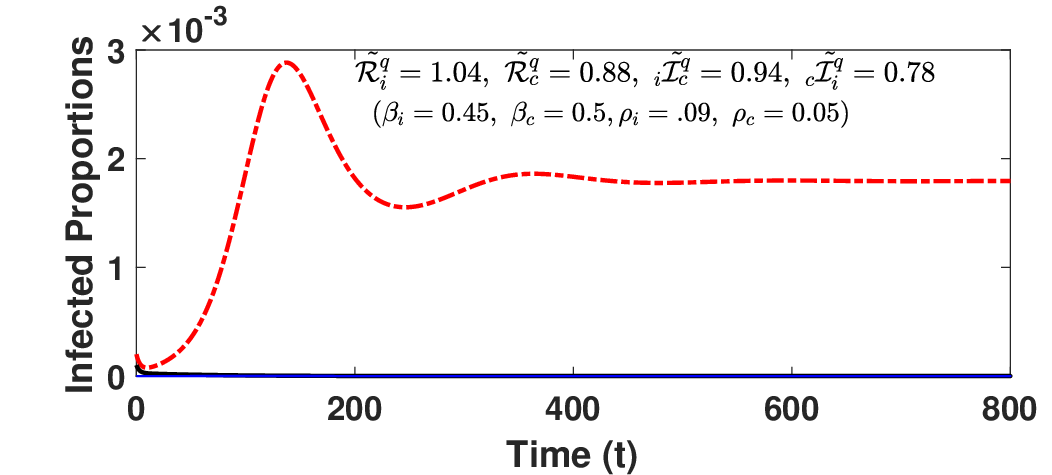}
  	\includegraphics[width=0.45\textwidth]{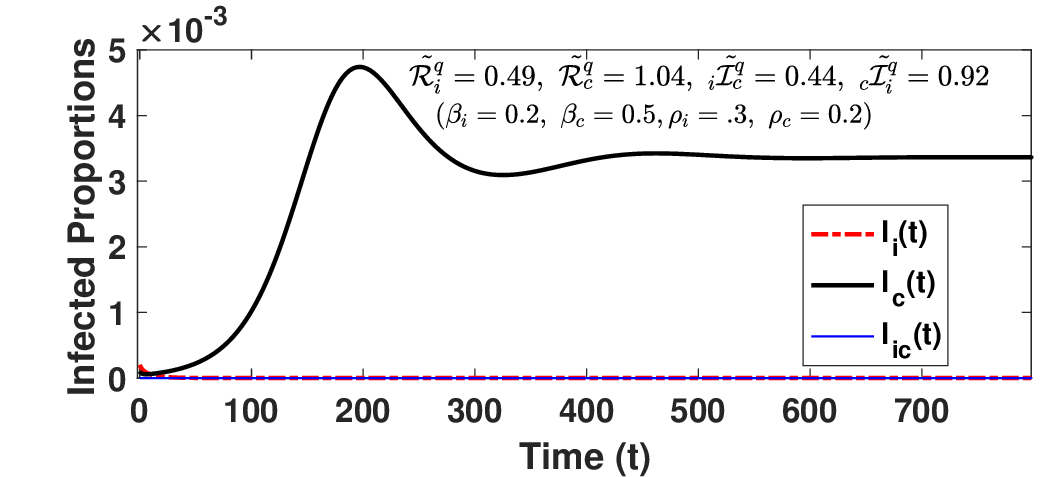}
    \includegraphics[width=.45\textwidth]{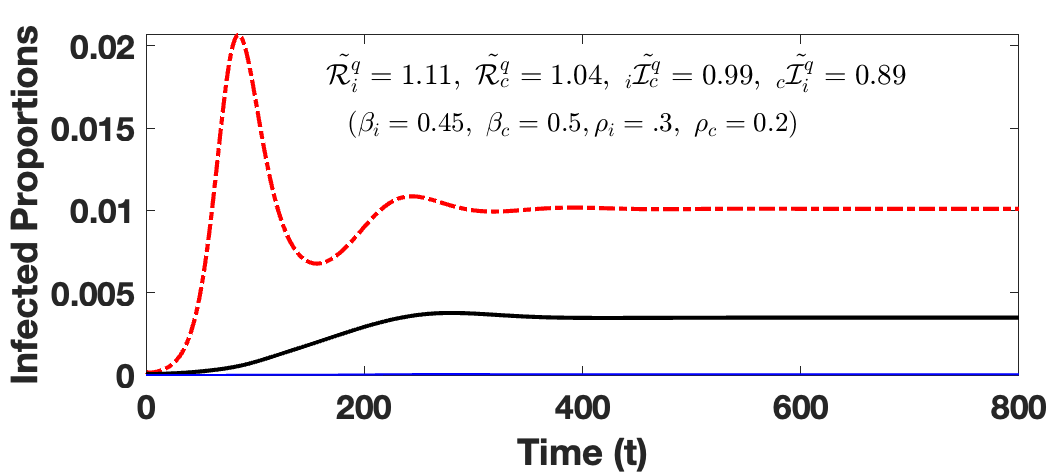}
  \includegraphics[width=.45\textwidth]{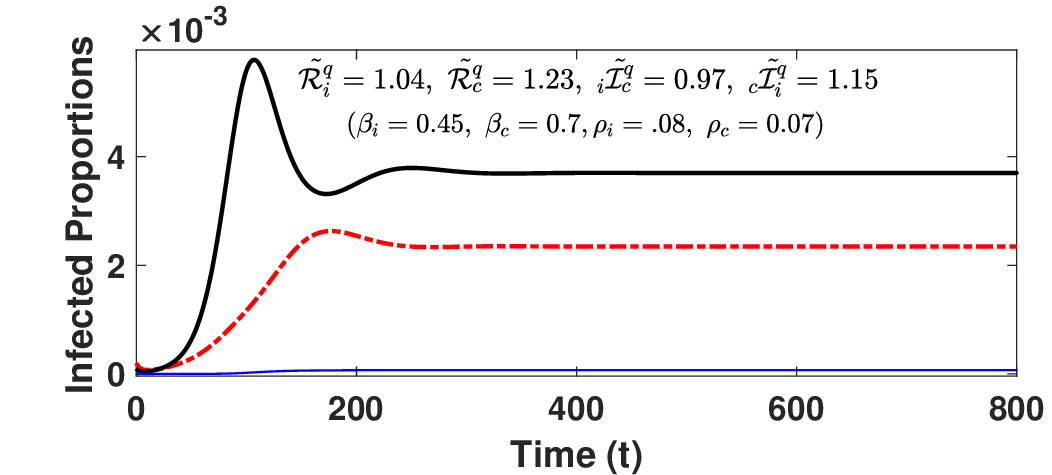}
\includegraphics[width=0.45\textwidth]{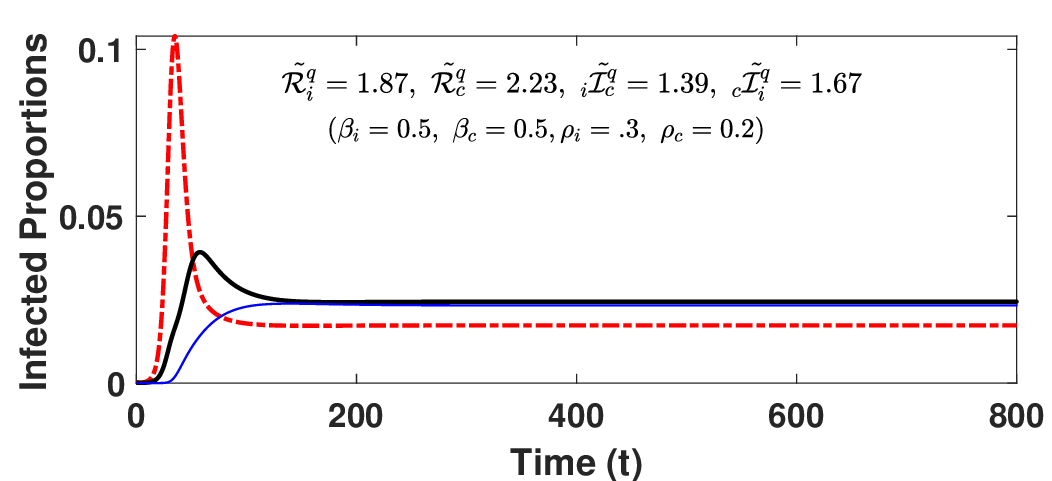}
  	\includegraphics[width=0.45\textwidth]{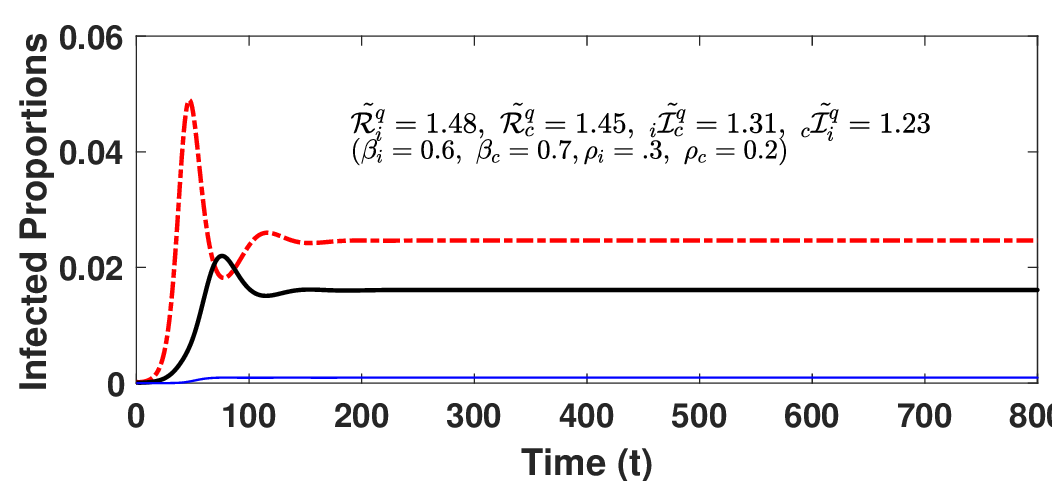}
\caption{Co-existence of multi-strain with co-infection.}
\label{fig:Coexistence_Imperfect_MainModel}
\end{figure}

Speaking broadly, model \cref{eqn:MainModel} is tedious to obtain analytical solutions. Therefore, in \cref{fig:Coexistence_Imperfect_MainModel}, we have shown numerically calculated solutions of model \cref{eqn:MainModel}  for given parameter values.  We found that the strain with the large reproduction number dominates the system. Moreover, whenever the invasion reproduction number(s) is (are) bigger than one, the co-infection strain remains in the population.

\subsection{Sensitivity and the quarantine impact analysis }
\subsubsection{Elasticity of Reproduction numbers} 

Primarily, sensitivity analysis of model parameters focuses on determining the most influential model parameters to plan the control strategies and to give guidance to the subsequent scientific work \cite{hamby1994review, samsuzzoha2013uncertainty}. In this subsection, we aim to investigate the relative importance of each model parameter included in the reproduction numbers using the differential sensitivity analysis as described in \cite{hamby1994review}. The sensitivity coefficient, $\Gamma_q$, of the independent variable, $q$, and dependent variable, $Q$,  obtained from the partial derivative of $Q$ to $q$. This result gives the percentage change in the quantity $Q$ to the percentage change in the parameter $q$ \cite{martcheva2015introduction}  is described by the formula:

\begin{align}
\Gamma_q = \dfrac{\partial Q}{\partial q}\times \dfrac{Q}{q}\approx \frac{\%\Delta Q}{\%\Delta q}
\end{align}

To assess how the model parameters impact the initial disease transmission, we performed a local sensitivity analysis in all parameters (independent variables) included in the reproduction numbers (dependent variables). The process was performed repeatedly by varying a single parameter at a time while holding the others constant. Computational outcomes are presented in \cref{tab:QRN_IRN_Sensitivity}.

\begin{table}[ht]
\small
	\centering
	\begin{tabular}{llllllllll}
		\toprule
{\multirow{3}{*}{\textbf{Symbol}}}	&	\multicolumn{9}{c}{\textbf{Sensitivity coefficient (\% change) } } \\
\cline{3-10}	
&&	\multicolumn{2}{c}{\textbf{QRN} }&&\multicolumn{2}{c}{\textbf{IRN} }&&\multicolumn{2}{c}{\textbf{CQRN} }\\
\cline{3-4}	\cline{6-7}	\cline{9-10}	
		 &&   $\pmb{\tilde{\mathcal{R}}_i^q}$& $\pmb{\tilde{\mathcal{R}}_c^q}$  && $\pmb{_i{\mathcal{I}^{q}_c}}$ & $\pmb{_c{\mathcal{I}^{q}_i}}$  && $\pmb{\mathfrak{R}_i^{c}}$& $\pmb{\mathfrak{R}_c^{i}}$ \\ \hline
		$\Lambda $ &&-&- &&$+4.5\times 10^{-10}$ &$+1.6\times 10^{-7}$&& -&- \\
		$\beta_{i } $ &&$+1.00000$&- && $+1.00000$&$-0.00113$&& $+1.00000$&- \\
		$\beta_{c} $&& -& $+1.00000$&&$-0.00118$ & $+1.00000$ &&- &$+1.00000$ \\
		$\rho_{i } $  && $+0.00932$ & -&& $+0.0565$& $-0.00014$&& -&- \\
		$\rho_{c } $&& - &$+032106$ && $-0.00827$& $+0.69968$&& -&- \\
		$\rho_{ic} $ && - & -&& $+1.3\times10^{-10}$& $+1.1\times10^{-6}$&& $+0.78563$ & $+0.78563$\\
		$\alpha_{i } $ && $-0.03753$& -&&$-0.05647$ & $+0.00016$&& -&- \\
		$\alpha_{c } $&& - &$-0.32102$ && $0.00827$& $-0.69964$&& -& -\\
		$\alpha_{ic} $ &&- & -&& $-1.3\times10^{-10}$& $-1.1\times10^{-6}$ &&$-0.78561$ &$-0.78561$  \\
		$\sigma_{i } $&& $-0.06626$ &- && $-0.10778$& $+0.00030$&&- &- \\
		$\sigma_{c } $ && -& $-0.03933$& &$+0.00101$&$-0.08576$ &&- & -\\
		$\sigma_{ic} $ &&- & -&& $-1.5\times10^{-10}$& $-1.3\times10^{-6}$  &&$-0.89861$ & $-0.89861$ \\
		$\tau_{i} $ && $+0.0120$ &- && $+0.01527$&$-0.00004$ && -&- \\
		$\tau_{c} $ && - & $+0.01894$&& $-0.00049$&$+0.04173$ && -& -\\
		$\tau_{ic} $ & &- &- && $+2.3\times10^{-11}$& $+2.0\times10^{-7}$  &&$+0.13437$ &$+0.13437$ \\
     	$\phi_{i } $ && $-0.05116$ &- &&$-0.01527$ &$+0.00004$ &&- &- \\
		$\phi_{c} $ && - & $-0.42200$&& $+0.00049$&$-0.04173$ && -&- \\
		$\phi_{ic } $ & &- &- && $-2.3\times10^{-11}$& $-2.0\times10^{-7}$  &&$-0.13437$ & $-0.13437$ \\
	    $\gamma_{i } $&&$-0.86599$ & -&&$-0.88922$ &$+0.00252$ && -&- \\
 		$\gamma_{c} $ &&- & $-0.52666$ &&$+0.01029$ &$-0.87066$&&- &- \\
		$\gamma_{ic} $ && -& -&& $-1.7\times10^{-11}$& $-1.5\times10^{-7}$  && $-10127$&$-10127$ \\
		$\delta_{i } $ &&$-0.00297$ &- && $-0.0029$& $+0.00167$&&- &- \\
		$\delta_{c} $&& - &$-0.0309$&&$+0.00051$ & $-0.04353$&& -&- \\
	   $\delta_{ic } $&&-  &- && $-1.7\times10^{-14}$& $-1.5\times10^{-10}$ && $-0.00010$&$-0.00010$ \\
	   $\mu $&& $-0.00011$ &$-0.00006$ && $-0.00006$& $-0.00011$&& $-0.00003$&$-0.00003$ \\
		\bottomrule
	\end{tabular}
\caption{Sensitivity indices of the quarantine reproduction numbers, QRN \& CQRN, and the invasion reproduction number(IRN). From the table, $+$ means a positive relationship; $-$means a negative relationship; - means no relationship. Parameter values are $\Lambda=0.0001$, $\beta_{i}=0.9$, $\beta_{c}=0.85$, $\rho_{i}=0.3$, $\rho_{c}=0.07$, $\rho_{ic}=0.05$, $\tau_{i}=0.02$, $\tau_{c}=0.014$, $\tau_{ic}=0.10$, $\alpha_{i}=0.02$, $\alpha_{c}=0.2$, $\alpha_{ic}=0.5$, $\sigma_{i}=0.04$, $\sigma_{c }=0.02$, $\sigma_{ic}=0.9$, $\delta_i = 0.001$, $\delta_c =  0.01$, $\delta_{ic} = 0.0001$, $\phi_i =  0.6$, $\phi_c = 0.92$, $\phi_{ic} = 0.7$, $\mu =  0.00001$, $\gamma_i = 0.3$, $\gamma_c = 0.2$, $\gamma_{ic} = 0.1$}
\label{tab:QRN_IRN_Sensitivity}
\end{table}

\subsubsection{Quarantine impact}
We analyze the efficacy of quarantine at the initial disease transmission to determine whether or not the increase of people in the quarantine facilities can be more beneficial or useless during the epidemic. It is evident  from \cref{tab:QRN_IRN_Sensitivity} that the elasticity of the quantity  $\tilde{\mathcal{R}}_k^q$ to its parameters is a monotonic decreasing function of  $\alpha_k$ and $\sigma_k$. With that in mind, \textit{we assess how the $\tilde{\mathcal{R}}_k^q$ will behave for the small and large values of these parameters}. 
For  $\sigma_k=0$ we evaluate  the critical level of the quarantine reproduction number corresponding to the limit $\alpha_k\rightarrow 0 $ and  $\alpha_k\rightarrow\infty$, respectively, we have

\begin{equation}
	\begin{array}{lll} 
		\tilde{\mathcal{R}^q_k}(\alpha_k^0)=\lim\limits_{\alpha_k\rightarrow 0} \tilde{\mathcal{R}_k^q} = \dfrac{\mathbf{r}_k\beta_k\rho_k}{\dot{\mathbf{p}}_k{\dot{\mathbf{q}}_k\mathbf{r}_k}}, &\text{ and,} & 			\tilde{\mathcal{R}^q_k}(\alpha_k^{\infty})=\lim\limits_{\alpha_k\rightarrow \infty} \tilde{\mathcal{R}_k^q} =\dfrac{\beta_k\mathbf{q}_k + \beta_k\tau_k}{\dot{\mathbf{q}}_k\mathbf{r}_k},
		\label{eqn:QRN_Alpha_limit}
	\end{array}
\end{equation}
where  $k=c \text{ or } i$.  The first and the second Equations of \cref{eqn:QRN_Alpha_limit}, correspondingly, give the maximum and the minimum number of secondary cases generated when only exposed are quarantined.
 On the other hand, for a fixed $\alpha_k=0$ the critical levels of the QRN corresponds to the limit $ \sigma_k\to 0$, and  $ \sigma_k\rightarrow\infty$, respectively, are given by: 
 
\begin{equation}
	\begin{aligned}
\tilde{\mathcal{R}^q_k}(\sigma_k^0)= \lim\limits_{\sigma_k\rightarrow 0} \tilde{\mathcal{R}^q_k} = \dfrac{\mathbf{r}_k\beta_k\rho_k}{\mathbf{p}_k\dot{\mathbf{q}}_k\mathbf{r}_k},~~~~ \text{ and }~~~~
	\tilde{\mathcal{R}^q_k}(\sigma_k^{\infty})=\lim\limits_{\sigma_k\rightarrow \infty} \tilde{\mathcal{R}^q_k} = \dfrac{\beta_k\rho_k}{\mathbf{p}_k\tilde{\mathbf{r}}_k},
	\end{aligned}
    \label{eqn:QRN_Sigma_limit}
\end{equation}
where  $k=c \text{ or } i$. The two equations given by \cref{eqn:QRN_Sigma_limit}, respectively, correspond to a maximum and minimum number of secondary infections generated when only infected are isolated. 
If none of the individuals is restricted or quarantined, in this case by setting $\alpha_k = \sigma_k = 0$, the that the corresponding reproduction number becomes
\begin{equation*}
	{\mathcal{R}}=\max\{{\mathcal{R}_c}, {\mathcal{R}_i}\}, ~\text{ where } {\mathcal{R}_k}= \dfrac{\beta_k\rho_k}{(\rho_k+\mu)( \gamma_k + \delta_k+\mu)} \text{~~and~~} k=c \text{ or } i
 \label{eqn:QRN_Alpha_Sigma_limit}
\end{equation*}

\begin{figure}[htbp]
  \centering
	\includegraphics[width=0.45\textwidth]{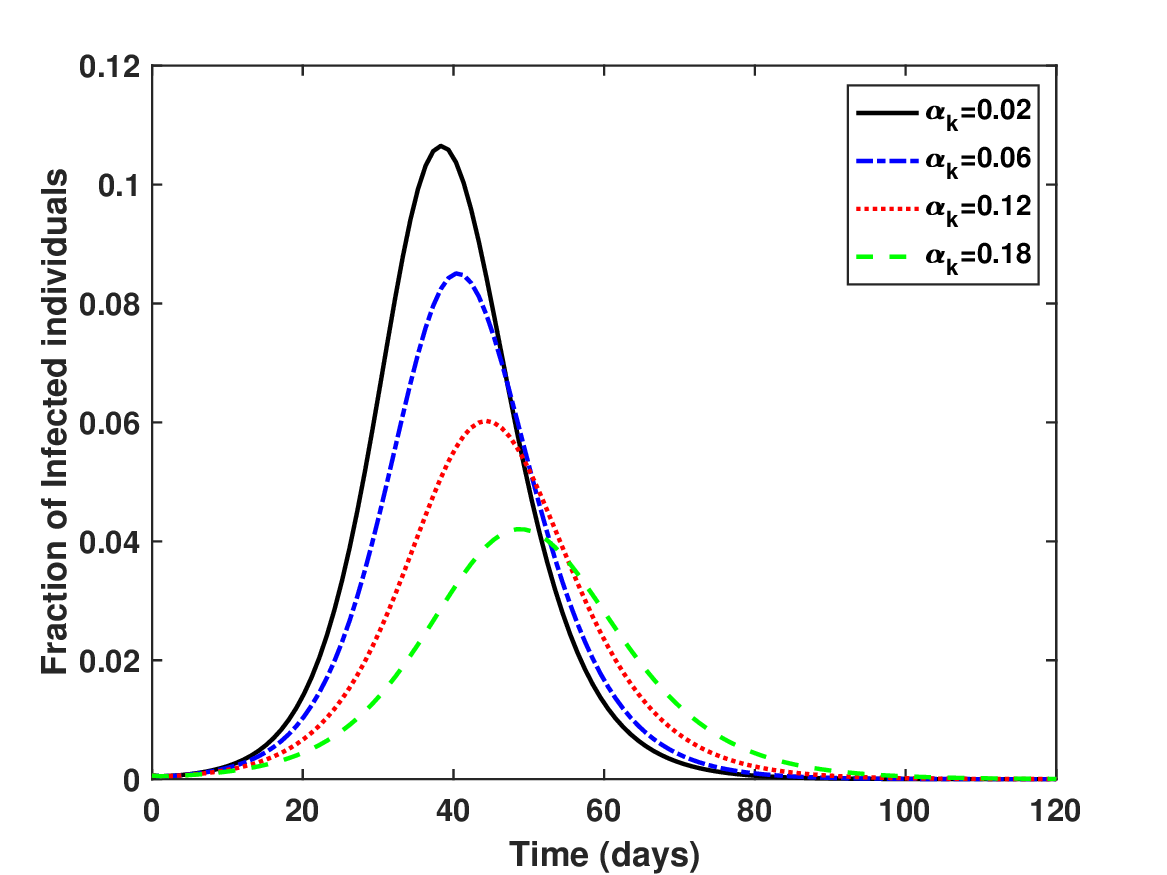}
  	\includegraphics[width=0.45\textwidth]{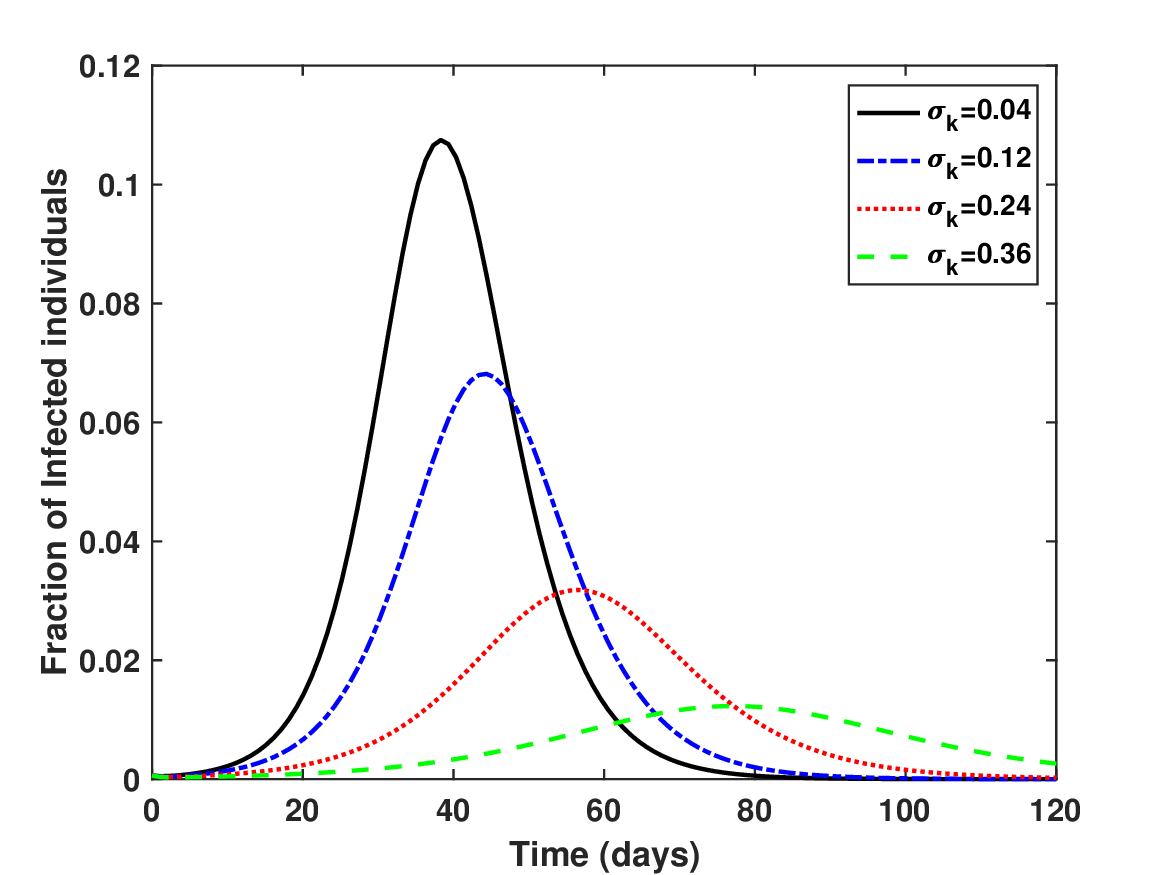}
    \includegraphics[width=.45\textwidth]{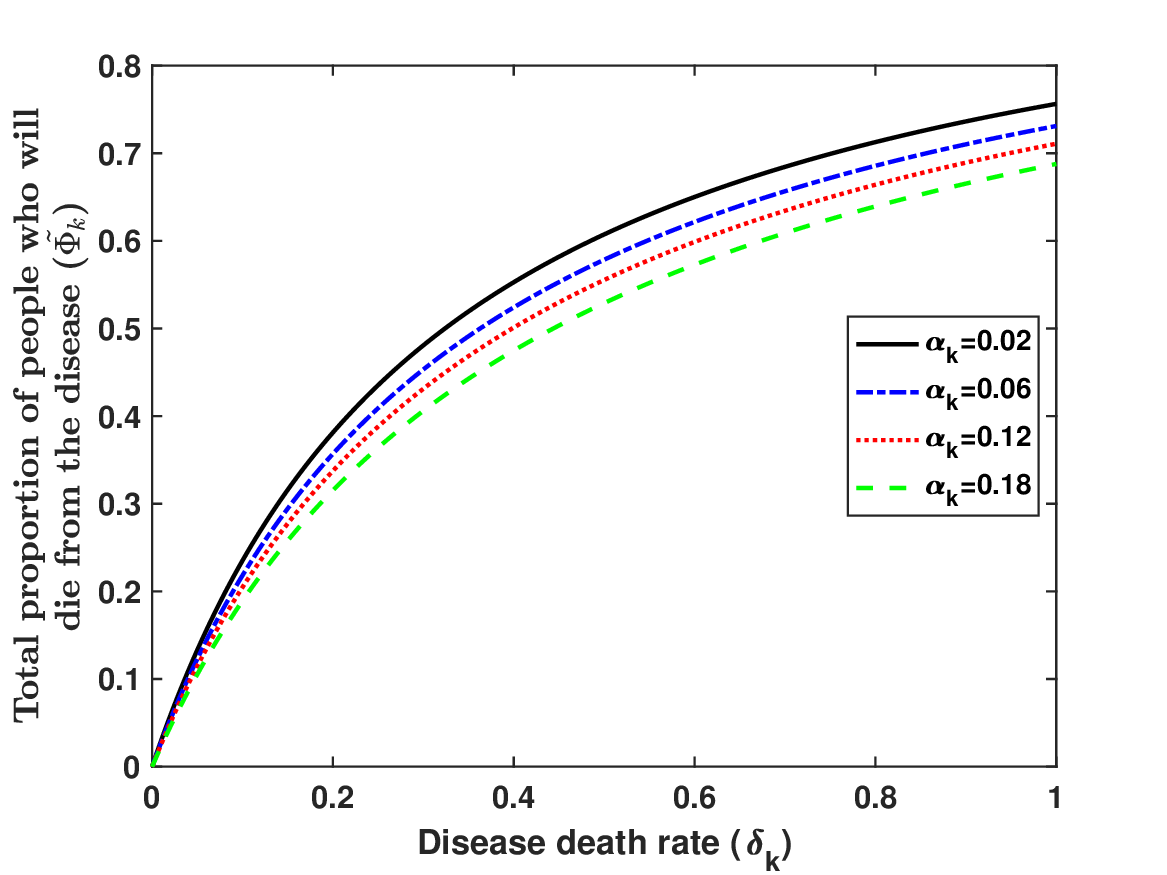}
    \includegraphics[width=.45\textwidth]{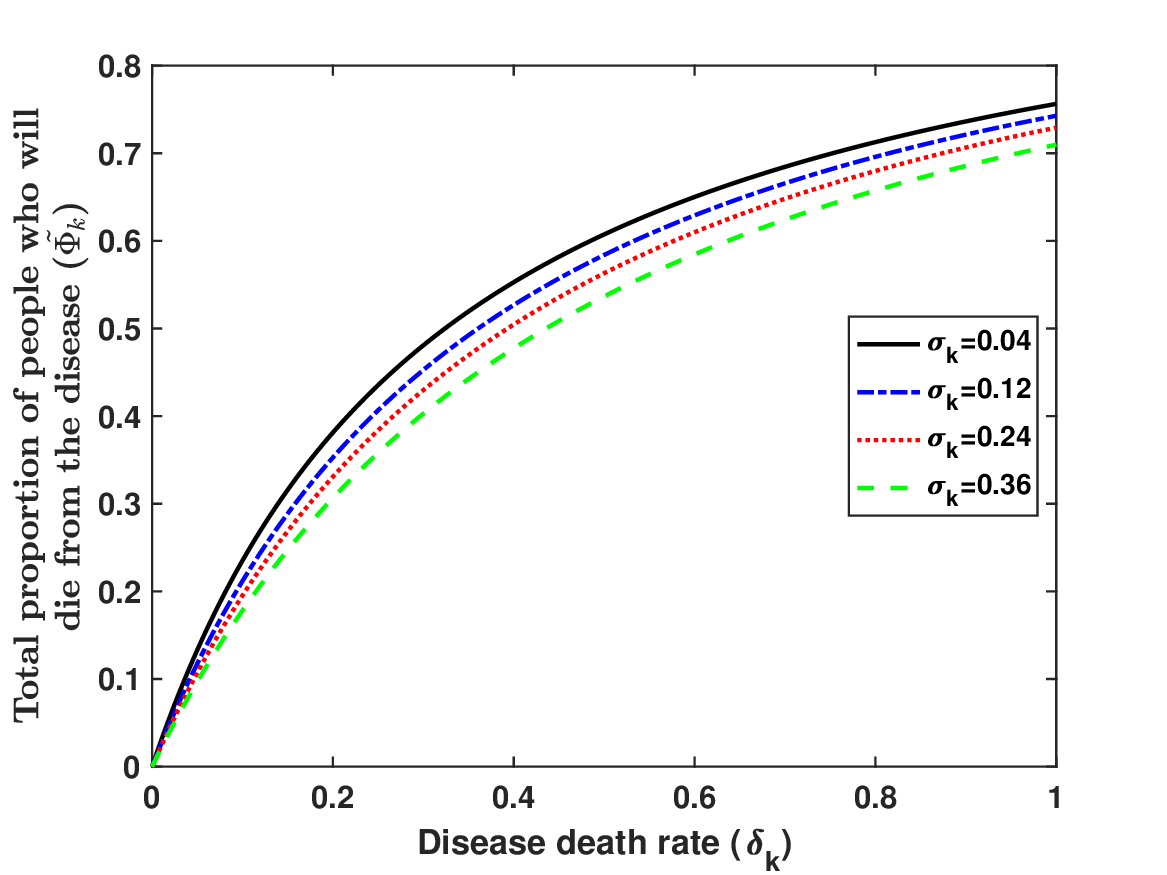}
	\caption{Numerical solutions of the fractions of infected individuals and the cumulative number of people who will die from the disease when $\alpha_k$ and $\sigma_k$ change. Note in top and bottom left figures $\sigma_k=0$. Further in top and bottom left figures $\alpha_k=0$. The remaining used parameters are $\Lambda=0.0001$, $\rho_k=0.3$, $\tau_k=0.06$, $\sigma_k=0.04$, $\phi_k=0.6$, $\mu=0.0001$, $\gamma_k=0.3$, $\beta_k=0.9$, $\omega_k=0.001$.}
	\label{fig:Consequences_reduced}
\end{figure}

\Cref{eqn:QRN_Alpha_limit} and \cref{eqn:QRN_Sigma_limit} implies that using quarantine (imperfect case) during an epidemic will not completely eradicate the disease in the population but will reduce the number of secondary infections and the disease complications, will decline accordingly. See \cref{fig:Consequences_reduced}. 

\subsubsection{Quarantine efficacy at the QRN }
To know the critical rates corresponding to the imperfect quarantine, we solve  for $\alpha_k$ and  $\sigma_k$ at $\tilde{\mathcal{R}_k^q}=1$ to obtain  $\alpha_k^*$ and  $\sigma_k^*$, respectively; they are given by

\begin{equation}
\begin{array}{lll}
\alpha_k^*=\dfrac{(\tilde{\mathcal{R}^q_k}(\alpha_k^0)-1)\dot{\mathbf{p}}_k}{1-	\tilde{\mathcal{R}^q_k}(\alpha_k^{\infty})} & \text{ and, } & \sigma_k^*=\dfrac{(\tilde{\mathcal{R}^q_k}(\sigma_k^0)-1)\dot{\mathbf{q}}_k}{1-\tilde{\mathcal{R}^q_k}(\sigma_k^{\infty})}.
\end{array}
\label{eqn:Alpha_Sigma_threshold}
\end{equation}

\begin{figure}[htbp]
  \centering
	\includegraphics[width=0.45\textwidth]{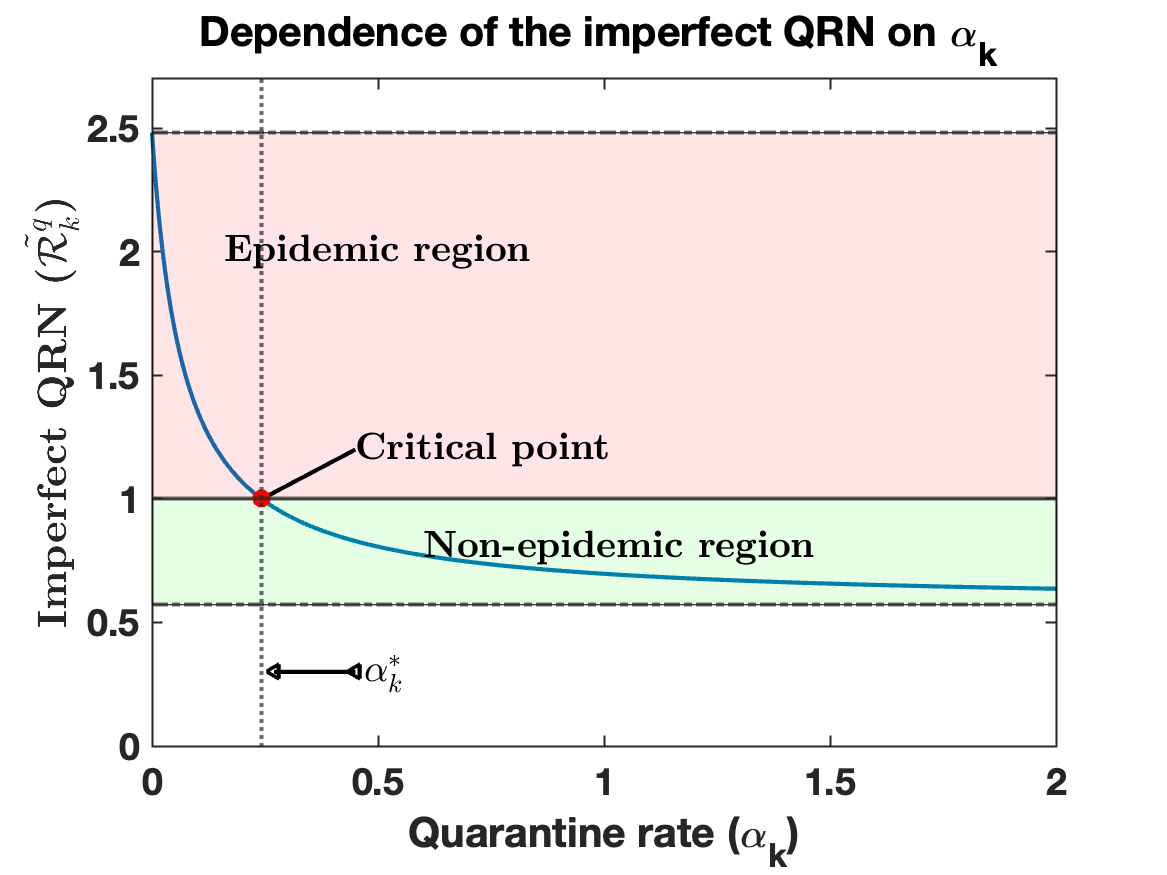}
  	\includegraphics[width=0.45\textwidth]{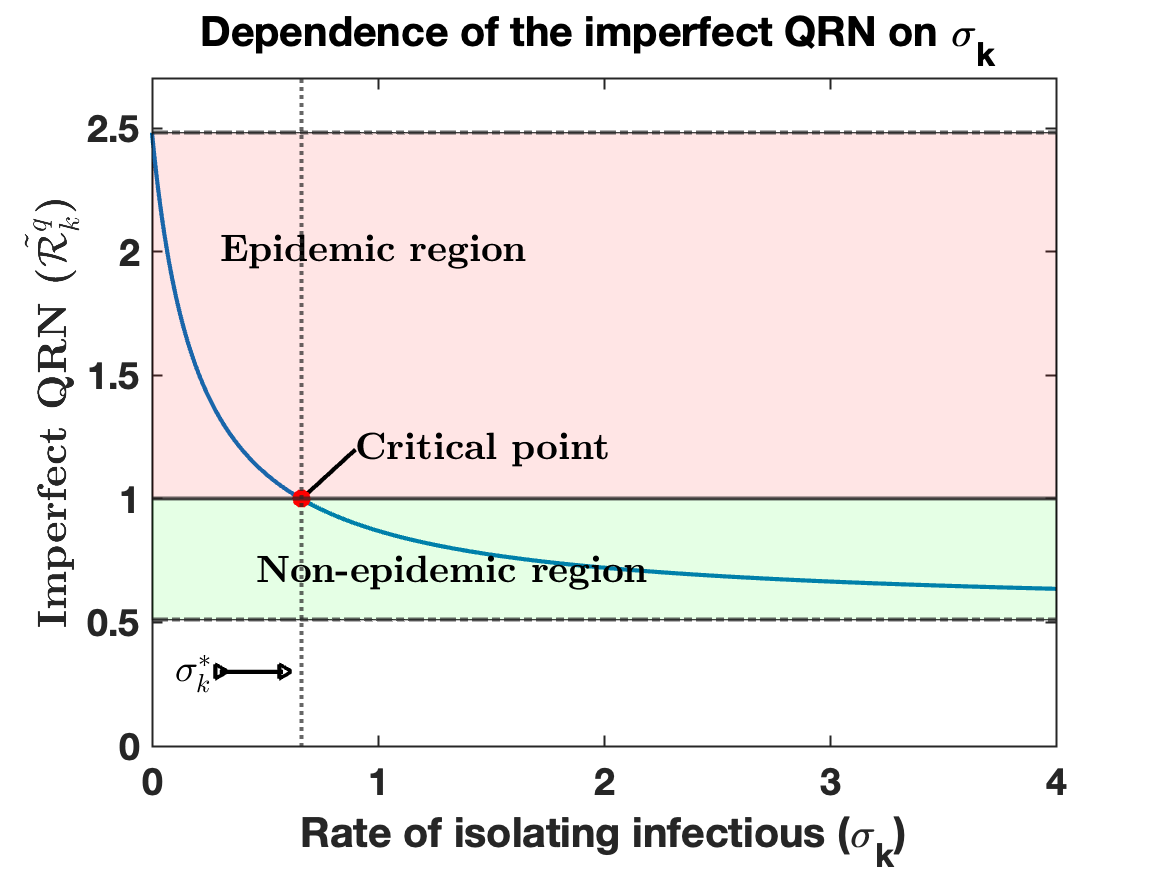}
    \includegraphics[width=.45\textwidth]{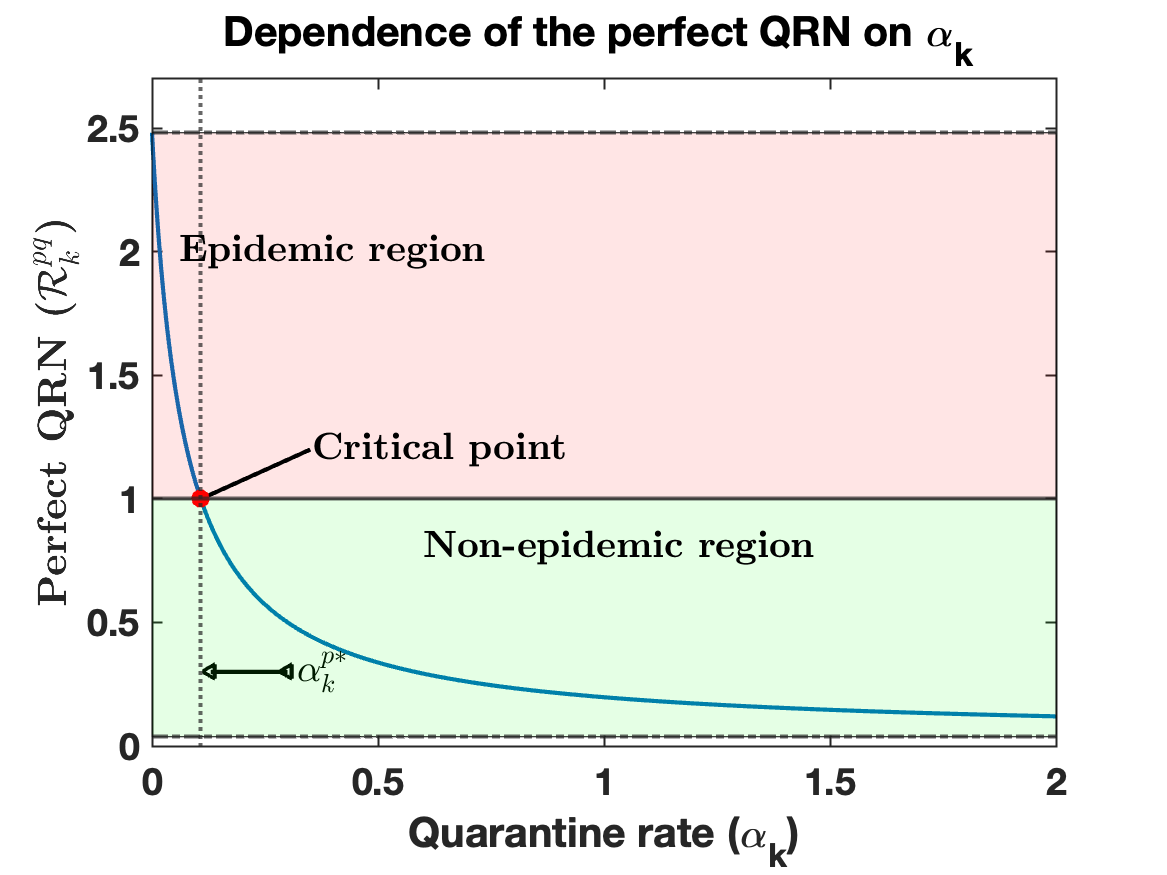}
    \includegraphics[width=.45\textwidth]{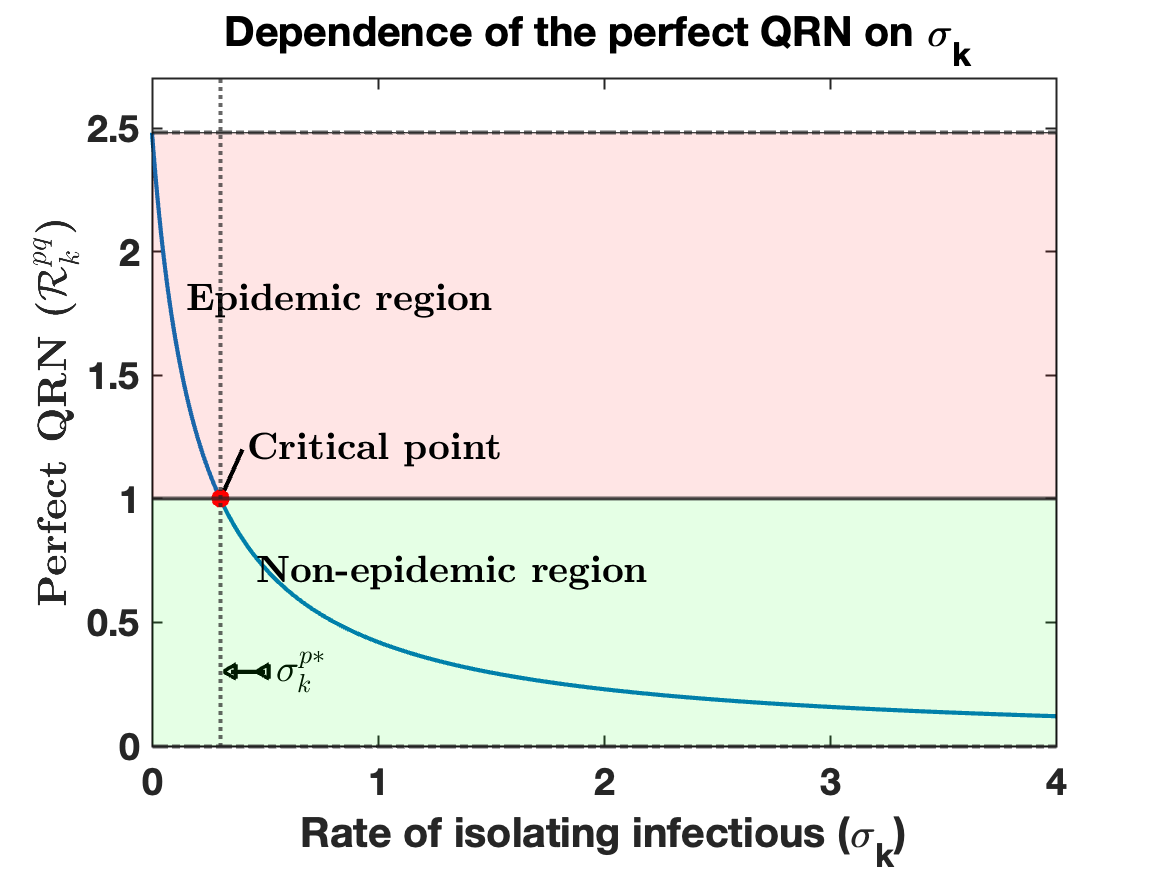}
	\caption{The dependence of the quarantine reproduction numbers (QRN) on the parameters related to the quarantine of exposed and infected, $\alpha_k$ and $\sigma_k$ respectively. The parameters are $\beta_k=0.5$, $\rho_k=0.07$, $\phi_k=0.92$, $\tau_k=0.015$, $\mu=0.0001$, $\gamma=0.2$, $\delta_k=0.001$.}
 \label{fig:Alpha_Sigma_QRN_dependences}
\end{figure}

For further analysis, we use equation \cref{eqn:DecreasingFunction} after replacing the variable $I_k$ by the expression of $\tilde{\mathcal{R}}^q_k$. The goal is to analyze how changes in $\alpha_k$ and $\sigma_k$ affect the quarantine efficacy at the initial disease transmission. Thus, 
\begin{align*}
	\Omega_k=1-\omega_k\tilde{\mathcal{R}}^q_k, 
\end{align*}
  where $\omega_k\in(0,1]$ (imperfect quarantine case). Differentiating $ \Omega_k$ with respect to $\alpha_k$ and evaluating the result at $\omega_k=1$, we obtain;
\begin{align}
	\dfrac{\partial \Omega_k}{\alpha_k}|_{\omega_k=1}=\dfrac{\tilde{\mathcal{R}}^q_k-\tilde{\mathcal{R}}^q_k(\alpha_k^{\infty})}{\mathbf{p}_c}>0 \text{positive impact if } \tilde{\mathcal{R}}^q_k>\tilde{\mathcal{R}}^q_k(\alpha_k^{\infty})
	\label{eqn:Efficacy_Exposed}
\end{align}
The above-obtained equation means that when only exposed individuals are quarantined, the imperfect quarantine will have a positive impact provided that $ \tilde{\mathcal{R}}^q_k>\tilde{\mathcal{R}}^q_k(\alpha_k^{\infty})$.
Similarly, differentiating $ \Omega_k$ with respect to $\sigma_k$ and evaluating the result at $\omega_k=1$, we have 
\begin{align}
	\dfrac{\partial \Omega_k}{\sigma_k}|_{\omega_k=1}=(\tilde{\mathcal{R}}^q_k-\tilde{\mathcal{R}}^q_k(\sigma_k^{\infty}))>0 \text{positive impact if } \tilde{\mathcal{R}}^q_k>\tilde{\mathcal{R}}^q_k(\sigma_k^{\infty}). \label{eqn:Efficacy_Infected}
\end{align}
The numerical simulation of the quarantine reproduction number dependences on quarantine-related parameters together with the critical values estimates given in Equations \cref{eqn:QRN_Alpha_limit} to	\cref{eqn:Efficacy_Infected} are visualized in \cref{fig:Alpha_Sigma_QRN_dependences}. Moreover, the figure compares the results of the perfect quarantine. With $\rho_k=0.07$, $\alpha_k=\sigma_k=[0,1]$, $\tau_k=0.014$, $\gamma=0.2$, $\phi_k=0.92$, $\beta_k=0.5$, $\delta_k=0.001$, and $\mu=0.0001$, the trajectories suggest that: the smaller the values of $\alpha_k$ and $\sigma_k$, the higher the value of threshold $\tilde{\mathcal{R}_k^q}$ and vice versa. The crucial fact to note is that implementing perfect quarantine at the initial transmission could effectively eradicate the disease. These results are consistent with the findings on optimal and sub-optimal control of the SARS model \cite{yan2008optimal}. 

\textit{Practicable questions, of course, are:  what fractions of individuals should be restricted if the quarantine is imperfect (i.e., $0<\Omega_k\leq 1$ ) with (i.) exposed only (ii.) infected only (iii.) both exposed and infected individuals?}
Let 
$$\alpha_k^q=\frac{\alpha_k}{\mathbf{r}_k},~~~~~ \text{ and, }~~~~~ \sigma_k^q=\frac{\sigma_k}{\mathbf{r}_k},$$ 
respectively, give the fractions of quarantined individuals from the exposed and the infected classes. We substitute $\alpha_k$ and $\sigma_k$ from \cref{eqn:Alpha_Sigma_threshold}  to obtain $\pmb{E}_k^{crit}$ and $\pmb{I}_k^{crit}$, the critical levels of the exposed and infected individuals need to be restricted when the quarantine facility is imperfect. These levels, respectively,  are given by
\begin{equation}
\begin{array}{lll}
	\pmb{E}_k^{crit}=\dfrac{(\tilde{\mathcal{R}}^q_k(\alpha_k^{0})-1)\dot{\mathbf{p}}_k}{(1-	\tilde{\mathcal{R}}^q_k(\alpha_k^{\infty}))\mathbf{r}_k}, &\text{ and, }&
\pmb{I}_k^{crit}=\dfrac{(\tilde{\mathcal{R}}^q_k(\sigma_k^{0})-1)\dot{\mathbf{q}}_k}{(1-	\tilde{\mathcal{R}}^q_k(\sigma_k^{\infty}))\mathbf{r}_k}.
	\label{eqn:Critical_levels}
\end{array}
\end{equation}

\begin{figure}[htbp]
  \centering
	\includegraphics[width=0.45\textwidth]{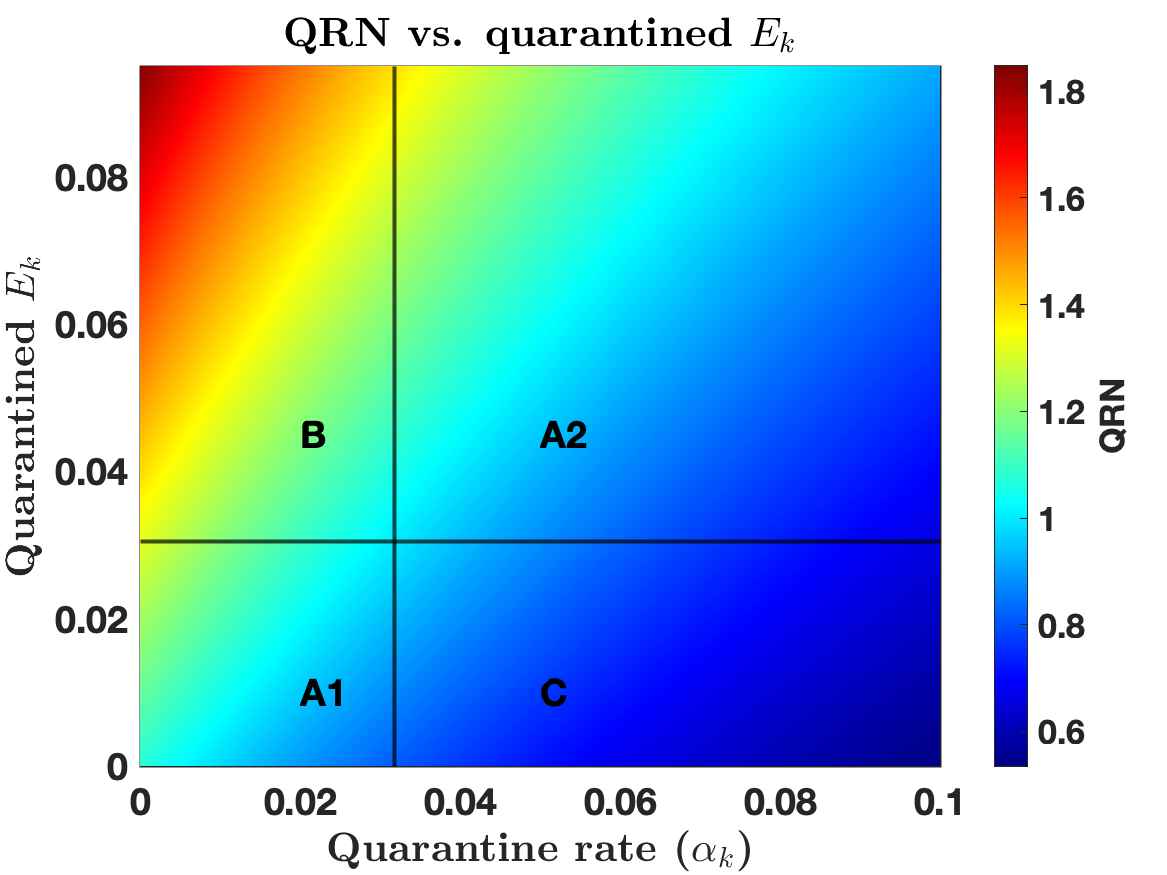}
  	\includegraphics[width=0.45\textwidth]{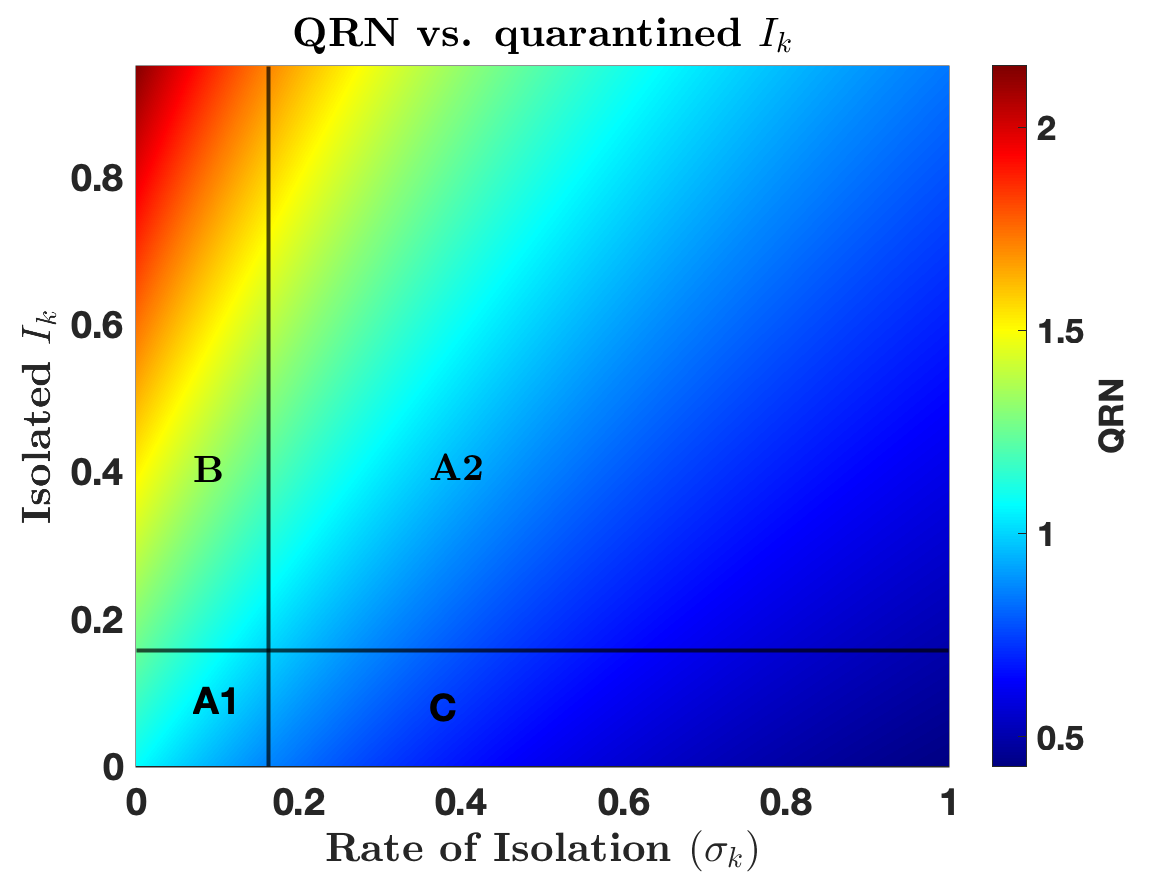}
	\caption{Relationship diagram showing how the parameters related to the quarantine (i.e., $\alpha_k$ and $\sigma_k$) and the proportions of individuals in the quarantine class influence the reproduction number. Here $\beta_k=0.78$, $\rho_k=0.09$, $\tau_k=0.0135$, $\delta_k=0.00001$, $\phi_k=1.02$,  $\mu=0.009$, and $\gamma_k=0.65$}
	\label{fig:Relationship_Alpha_Sigma_QRN}
\end{figure}

In \cref{fig:Relationship_Alpha_Sigma_QRN}, we estimate the fractions of quarantined individuals from the exposed and the infected classes and then determine the secondary cases for various $\alpha_k$ and $\sigma_k$. Moreover, we  indicates the critical levels, $\pmb{E}_k^{crit}$ and $\pmb{I}_k^{crit}$. The plots illustrate that applying quarantine (in the case of imperfect quarantine) will help to lower the secondary cases to less than unity if the rates of sending people to the quarantine are at least their critical values. In addition, the total number of people in the quarantine facility should not exceed its critical values (see region C  in \cref{fig:Relationship_Alpha_Sigma_QRN}. \Cref{lem:Quarantine_impact_QRN} summarizes these results.

\begin{lemma}
	Given  imperfect model \cref{eqn:MainModel_Reduced},
	\begin{enumerate}
		\item  when only exposed individuals are quarantined, the imperfect quarantine will have positive impact if and only if $\alpha_k\geq \alpha_k^*$, $\alpha_k^q\leq  \pmb{E}_k^{crit}$, and  $\tilde{\mathcal{R}}_q^{k}>\tilde{\mathcal{R}}^q_k(\alpha_k^{\infty})<1$. 
		\item when only infected individuals are isolated, the imperfect quarantine will have positive impact if and only if $\sigma_k\geq \sigma_k^*$, $\sigma_k^q\leq  \pmb{I}_k^{crit}$, and  $\tilde{\mathcal{R}}_q^{k}>\tilde{\mathcal{R}}^q_k(\sigma_k^{\infty})<1$; 
		\item when both exposed and infected individuals are restricted, we estimate the following results
		\begin{enumerate}
			\item the overall rate of sending people in the quarantine should be  $\max\{\alpha_k^*,\sigma_k^*\}$,
			\item the total number of people in the quarantine should be $\min\{\pmb{E}_k^{crit}, \pmb{I}_k^{crit}\}$.
		\end{enumerate}
	\end{enumerate}
	\label{lem:Quarantine_impact_QRN}
\end{lemma}

\begin{figure}[htbp]
  \centering
	\includegraphics[width=0.9\textwidth]{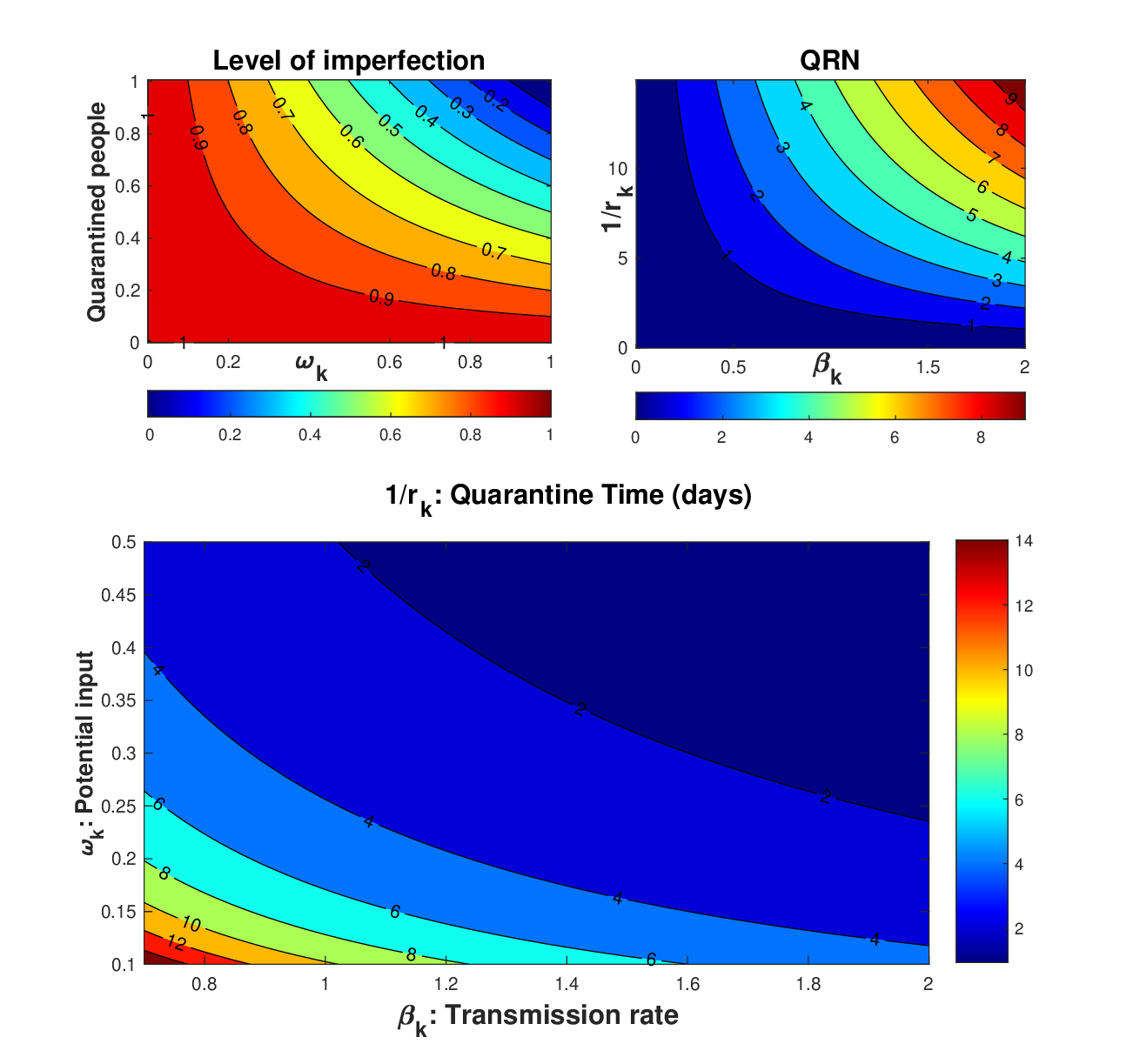}
	\caption{The \textbf{ top left figure} estimates change in the level of imperfection, $\Omega_k$, to the effort provided, $\omega_k$, and the proportion number of individuals quarantined. The expected time in the quarantine concerning the potential parameter $\omega_k$ and the transmission rate $\beta_k$ is given in the bottom figure. The \textbf{top right figure} gives the relationship between the transmission rate $\beta_k$, quarantine time, and infectiousness profile in terms of quarantine reproduction number, QRN. Parameters are taken as in \cref{fig:Relationship_Alpha_Sigma_QRN}. }
\label{fig:Effectiveness_Time_Infectiousness} 
\end{figure}

\Cref{fig:Effectiveness_Time_Infectiousness}  gives some justification for the level of imperfection concerning the number of people in the quarantine and the potential input provided to control the disease. On top of that, it estimates the dependencies of quarantine time to the parameter $\omega_k$. Lastly, the figure estimates the disease profile to the transmission rate and time to quarantine. 

\section{Conclusion and future work}\label{sec:conclusions}

We proposed the model with a single host population and multiple strains with co-infections. We started analyzing the model by assuming that only a single pathogen invaded the population. We examined the reproduction number and local stability using the next-generation approach. Employing the criterion introduced by Castillo-Chavez and his colleagues, we performed the global stability of the disease-free equilibrium. 

Using the computer algebra systems, numerically, we solved the multi-strains model to predict the complex endemic equilibrium solution of the system. The ($\ddot{\xi_{c}^*}, \ddot{\xi_{i}^*}$) relationship in \cref{fig:coexistence} suggests that several endemic solutions are possible depending on the parameter values. As in \cref{prop:DFE_MainModel_Reduced}, \cref{fig:Coexistence_Imperfect_MainModel} shows that if $\tilde{\mathcal{R}}_k^q>1$ (here $k=c \text{ or } i $) then the corresponding pathogen will dominate the system. Analysis in \cref{fig:CROSS1} tells how the cross-immunity parameters and the coexistence region are related. From a modeling point of view, this relationship notifies that; the smaller the value of the cross-immunity parameters, the smaller the fraction of the recovered population from one strain is susceptible to the second pathogen, consequently reducing the co-exposed individuals.

Moreover, our results provide a mathematical analysis of the impact of quarantine on the initial disease transmission and determine critical levels based on quarantine-related parameters. We note that the quarantine and isolation of exposed and infected individuals will reduce the number of secondary cases consequently, reduce disease complications. See, for
example, \cref{fig:Consequences_reduced}. Here we illustrated examples of disease complications: peak magnitude and people who will die from the disease. We found that as $\alpha_k$ and $\sigma_k$ increase, the peak magnitude and the cumulative number of people who will die from the disease decline over time. However, the peak time and the final size could increase significantly. 

Generally, proper application of rules and regulations (like waste management, accessibility to health care, and issues about hygiene) in the quarantine could result in less possibility of infection (less imperfection).  See the top left illustration of \cref{fig:Effectiveness_Time_Infectiousness}. Additionally, from the bottom figure, as $\omega_k \to 0$,  the quarantine duration is anticipated to be longer than when $\omega_k \to 1$, we can plausibly argue that the role played by the respective authority is essential to shorten the quarantine duration however, other parameters must be considered, for instance, the rates of transmission and recovery.

Here we made a few remarks. From a modeling point of view, the phenomenon of most co-infection diseases is still only partially understood. Principally, the intention of the proposed model was not to give a complete picture of the underlying behavior of competing or co-existing pathogens. Nevertheless, it provides the concept of various existing equilibria, the impact of quarantine as a control measure on the initial dynamic of the diseases, and cross-immune response. In addition,
we solved the model \cref{eqn:MainModel} numerically, some parameter values when not available, so the assumed values were within a realistic range for illustrative purposes. Lastly, the linear equation \cref{eqn:DecreasingFunction} proved to work better in a model that used data \cite{mitchell2016data}. However, the choice of function was reasonable to account for the decline of infections upon applying control measures, i.e., quarantine; one could use other types of declining functions.

\appendix
\section{Positivity and Boundedness} \label{lem:Postivity_Boundedness}
\subsection{Positivity of solutions of model \ref{eqn:MainModel}} \label{ssst:Positivity}
\begin{lemma}[Positivity of solution]\label{lem:Non-negativity} Let $$(S(t),E_i(t),E_c(t), E_{ic}(t) ,I_i(t), I_c(t), I_{ic}(t), Q_i(t), Q_c(t), Q_{ic}(t), R_i(t),R_c(t), R_{ic}(t))$$ be the solution of the system \cref{eqn:MainModel}. With initial conditions \cref{eqn:MainModel_InitialCon}, the solution exists and remains non-negative for all $t \geq 0$.
\end{lemma}

\begin{proof}
Consider system \cref{eqn:MainModel},	it follows from \cref{eqn:MainModel_InitialCon} that 
\begin{align*}
	\begin{array}{l}
	\dfrac{dS}{dt}|_{S=0}=\Lambda>0,	\\
        \dfrac{dE_{i}}{dt}|_{E_{i}=0}=\xi_{i}S+\eta_cR_c\xi_i\geq 0, \forall~ S,R_c\geq 0,	\dfrac{dE_{c}}{dt}|_{E_{c}=0}=\xi_{c}S +\eta_iR_i\xi_c, \forall~ S,R_i\geq 0,\\
		\dfrac{dE_{ic}}{dt}|_{E_{ic}=0}=\xi_{i}I_c +\xi_cI_i \geq 0, \forall~ I_i,I_c\geq 0,\dfrac{dI_{i}}{dt}|_{I_{i}=0}=\rho_iE_i+\tau_iQ_i  \forall~ E_i,Q_i\geq 0,\\
		\dfrac{dI_{c}}{dt}|_{I_{c}=0}=\rho_cE_c+\tau_cQ_c  \forall~ E_c,Q_c\geq 0,	\dfrac{dI_{ic}}{dt}|_{I_{ic}=0}=\rho_{ic}E_{ic}+\tau_{ic}Q_{ic}  \forall~ E_{ic},Q_{ic}\geq 0, \\
				\dfrac{dQ_{i}}{dt}|_{Q_{i}=0}=\alpha_iE_i+\sigma_iI_i~\forall~ E_i,I_i\geq 0,
		\dfrac{dQ_{c}}{dt}|_{Q_{c}=0}=\alpha_cE_c+\sigma_cI_c~\forall~ E_c,I_c\geq 0,\\
		\dfrac{dQ_{ic}}{dt}|_{Q_{ic}=0}=\alpha_{ic}E_{ic}+\sigma_{ic}I_{ic}~\forall~ E_{ic},I_{ic}\geq 0,
		\dfrac{dR_{i}}{dt}|_{R_{i}=0}=\phi_iQ_i+\gamma_iI_i~\forall~ Q_{i},I_{i}\geq 0,\\
		\dfrac{dR_{c}}{dt}|_{R_{c}=0}=\phi_cQ_c+\gamma_cI_c~\forall~ Q_{c},I_{c}\geq 0,
		\dfrac{dR_{ic}}{dt}|_{R_{ic}=0}=\phi_{ic}Q_{ic}+\gamma_{ic}I_{ic}~\forall~ Q_{ic},I_{ic}\geq 0.\\
		\end{array}
	\end{align*}
We now prove that all solutions of  (\ref{eqn:MainModel}) are non-negative as $t\to\infty$. From the first equation of the system, we have
		\begin{align*}
			\dfrac{dS}{dt}=\Lambda - (\xi_{i}+\xi_{c} + \mu) S\geq  - (\xi_{i}+\xi_{c} + \mu) S,
	\end{align*}
after integration we obtain
\begin{align*}
S(t)=S(0)e^{-\bigintsss_{0}^t (\xi_{i}+\xi_{c} + \mu)ds}> 0 \text{ for all } t>0.
\end{align*}
It can be shown that $E_i(t)>0,E_c(t)>0, E_{ic}(t)>0 ,I_i(t)>0, I_c(t)>0, I_{ic}(t)>0, Q_i(t)>0, Q_c(t)>0,$ $Q_{ic}(t)>0, R_i(t)>0,R_c(t)>0, R_{ic}(t)>0$ \text{ for all } $t>0$ analogously to $S(t)$.
\end{proof}

\subsection{Boundedness of solutions}\label{ssst:boundedness}
 \begin{lemma}[Boundedness of solution]\label{lem:Boundedness} 
	The closed set 
	\begin{align*}
		\mathcal{C}= &\Big\{ (S,E_i,E_c, E_{ic},I_i,I_c, I_{ic},Q_i,Q_c, Q_{ic},R_i,R_c, R_{ic})\in\mathbb{R}^{13}_+: \\ &S+E_i+E_c+ E_{ic}+I_i+I_c+I_{ic}+Q_i+Q_c+Q_{ic}+R_i+R_c+ R_{ic}\leq\dfrac{\Lambda}{\mu}\Big\}
	\end{align*}
	is positively-invariant for the model \cref{eqn:MainModel}.
\end{lemma}

\begin{proof}
		Adding all equations in the model (\ref{eqn:MainModel}) gives the rate of change in a total population over time:
		\begin{align}
			N'_T(t) = \Lambda - \mu N_T - (\delta_c I_c + \delta_c Q_c +\delta_iI_i + \delta_iQ_i +\delta_{ic}I_{ic} + \delta_{ic}Q_{ic}).
			\label{eqn:Overall}
		\end{align} 
		From the equation (\ref{eqn:Overall}), we have the condition \begin{align}
			N'_T(t) \leq \Lambda - \mu N_T,
			\label{eqn:Overall_}
		\end{align} 
		satisfied. Thus, the total population is bounded above, $	N'_T(t)\leq 0$ whenever $N_T(t)\geq \Lambda/\mu$. From the equation (\ref{eqn:Overall_}),  it is easy to see that
		$$N_T(t)\leq\frac{\Lambda}{\mu}+\left(N_T(0)-\frac{\Lambda}{\mu}\right)e^{-\mu t}.$$

		For any $t\geq 0$, the inequality above, gives
		\begin{align}
			0<N_T(t)\leq
			\begin{cases}
				N_T(0), & \text{ for  } ~~t=0,\\
				\Lambda/\mu,& \text{ as }~~ t\to\infty\\
			\end{cases}
			\label{eqn:InvariantFunction}
		\end{align}
		Hence, the region $\mathcal{C}$ is positive invariant and attracts all solutions in $\mathbb{R}^{13}_+$. 
	\end{proof}
	\noindent We have shown in \cref{lem:Boundedness} that region $\mathcal{C}$ is positive-invariant and attractive. Thus, it is sufficient to investigate the dynamics of \cref{eqn:MainModel} in $\mathcal{C}$, where the unique solution exists and continuously depends on the system data. 
 
\subsection{Non-negativity and boundness of solution of \ref{eqn:MainModel_Reduced} }\label{app:Postivity_Boundedness_Model_Reduced}
\begin{proposition} Let $(S(t),E_k(t),I_k(t), Q_k(t), R_k(t) )$ be the solution of the system \cref{eqn:MainModel_Reduced}.
		\begin{enumerate}
			\item With initial conditions \cref{eqn:MainModel_InitialCon}, the solution exists and remains non-negative for all $t \geq 0$.
			\item   The closed set 
			\begin{align*}
				\mathcal{C}_k= &\Big\{ (S,E_k,I_k,Q_k,R_k)\in\mathbb{R}^{5}_+: S+E_k+I_k+Q_k+R_k\leq\dfrac{\Lambda}{\mu}\Big\}
			\end{align*}
			is positively-invariant for the model (\ref{eqn:MainModel_Reduced}).
		\end{enumerate}
		\label{prop:Boundedness_Reduced}
	\end{proposition}
	\begin{proof}
	 A similar way used to prove the positivity and boundedness of system \cref{eqn:MainModel} (i.e. \cref{ssst:basic_properties}) can be used to show that the solutions of system \cref{eqn:MainModel_Reduced} start in the set $\mathcal{C}_k$ and remain there for all $t > 0$ and that $\mathcal{C}_k$ attracts all solutions in $\mathbb{R}^{5}_+$. Therefore, it is satisfactory to recognize the dynamics of \cref{eqn:MainModel_Reduced} in $\mathcal{C}_k$.
	\end{proof}
 
\section{Global Stability of the DFE-(\texorpdfstring{$\ddot{\mathbf{E}}^0$)}{Lg}}\label{sec:GAS_Coex}
We start by considering the closed set 	\begin{align*}
	\mathfrak{C}= &\Big\{ (S,E_i,E_c,I_i,I_c,Q_i,Q_c,R_i,R_c)\in\mathbb{R}^{9}_+: \\ &S+E_i+E_c+I_i+I_c+Q_i+Q_c+R_i+R_c\leq\dfrac{\Lambda}{\mu}\Big\}
\end{align*}
\begin{lemma} The region $	\mathfrak{C}$ is positively-invariant for the model (\ref{eqn:MainModel_Reduced2}).
	\label{lem:Boundedness_Reduced2} 
\end{lemma}
\begin{proof}
In \cref{lem:Boundedness}, we have shown that the region $\mathcal{C}$ is positively invariant and attractive. Since  $\mathfrak{C}\subset\mathcal{C}$ by assumption, the region $\mathfrak{C}$ is positively invariant and attractive. Thus, we recognize the coexistence model (\ref{eqn:MainModel_Reduced2}) to make biological sense in $\mathfrak{C}$.
\end{proof}
Next, write system \cref{eqn:MainModel_Reduced2} in the form of equation 3.1 in \cite{castillo2002computation} as follows: we denote a vector of uninfected individuals by $ \mathbf{x}=(S, R_i,R_c)^T\in\mathbb{R}^3$ and that of infected individuals by $ \mathbf{I}=(E_i, E_c, I_i,I_c, Q_i,Q_c)^T\in\mathbb{R}^6$ such that 
\begin{equation}
	\begin{aligned}
		\dfrac{d\mathbf{x}}{dt}=F(\mathbf{x}, \mathbf{I}),~~
		\dfrac{d\mathbf{I}}{dt}=G(\mathbf{x}, \mathbf{I}). 
		\label{eqn:MSM_Global}
	\end{aligned}
\end{equation}
We also denote  the DFE, $\ddot{\mathbf{E}}^0=(\mathbf{x}^*,0,0,0,0,0,0)$  where $\mathbf{x}^*$ define the DFE of system ${d\mathbf{x}}/{dt}$. Moreover, we state the following conditions:
\begin{itemize}
	\item [\textbf{H1:}] For $F(\mathbf{x}, \mathbf{I})|_{\mathbf{x}^*}$, $\mathbf{x}^*$ is globally asymptotically stable  $GAS$,
	\item [\textbf{H2:}] $G(\mathbf{x}, \mathbf{I})=A\mathbf{I}-\hat{G}(\mathbf{x}, \mathbf{I})$, $\hat{G}(\mathbf{x}, \mathbf{I})\geq 0 $ for $(\mathbf{x}, \mathbf{I})\in\mathfrak{C}$,
\end{itemize}

\noindent where $A=G(\mathbf{x}^*, 0)$ is an $M$-matrix. The following claim is true when the given two conditions hold for the system \cref{eqn:MainModel_Reduced2}:

 \begin{theorem}
	The DFE  $\ddot{\mathbf{E}}^0=(\mathbf{x}^*,0,0,0,0,0,0)$ of system (\ref{eqn:MSM_Global}), equivalent to (\ref{eqn:MainModel_Reduced2})  is globally asymptotically stable ($GAS$) if assumptions \textbf{H1} and \textbf{H2} holds and that \\$\max\{\tilde{\mathcal{R}^q_c},\tilde{\mathcal{R}^q_i} \}< 1$ ($LAS$).
\end{theorem}
\begin{proof} Using next-generation approach we have shown in \cref{ssst:DFE_MSM} that $$\max\{\tilde{\mathcal{R}^q_c},\tilde{\mathcal{R}^q_i} \}< 1~~(LAS).$$  Let us now consider the prove for assumptions \textbf{H1} and \textbf{H2}.  From (\ref{eqn:MSM_Global}), we have 
	\begin{equation*}
		\begin{array}{l}
			F(\mathbf{x}, 0)= \begin{pmatrix}
				\Lambda -\mu S\\
				0 \\
				0
			\end{pmatrix}, ~~ A= \begin{pmatrix}
				-\mathbf{p}_1 & 0 & \beta_i & 0& \beta_i&0\\ 
				0& -\mathbf{p}_2 & 0 & \beta_c & 0& \beta_c\\ 
				\rho_i&0 & -\mathbf{q}_1 &0& \tau_i&0\\
				0&	\rho_c&0 & -\mathbf{q}_2 &0& \tau_c\\
				\alpha_{i} &0&\sigma_{i}& 0&-\mathbf{r}_1&0\\
				0&	\alpha_{c} &0&\sigma_{c}& 0&-\mathbf{r}_2\\
			\end{pmatrix},   \\
			\text{ and, } \\
			\hat{G}(\mathbf{x}, \mathbf{I})=\begin{pmatrix}
				\beta_iI_i(1-\frac{S}{N})+\beta_iQ_i(1-\frac{\Omega_{i}S}{N})\\
				\beta_cI_c(1-\frac{S}{N})+\beta_cQ_c(1-\frac{\Omega_{c}S}{N})\\
				0\\
				0\\
				0\\
				0
			\end{pmatrix}.
		\end{array}
	\end{equation*}
	
\noindent	It follows that $\Omega_i,\Omega_{c}\in(0,1]$,  and $0\leq \Omega_i S,\Omega_{c} S\leq S\leq N$ then $  	\hat{G}(\mathbf{x}, \mathbf{I})\geq 0$ (\textbf{H2} holds).  Moreover,
	\begin{align*}
		\lim\limits_{t\to\infty}F(\mathbf{x}(t), \mathbf{I}(t))|_{\mathbf{x}^*}=\lim\limits_{t\to\infty}F(\mathbf{x}(t), 0)=(\frac{\Lambda}{\mu}, 0,0)=\mathbf{x}^*, \text{\textbf{ H1} holds}.
	\end{align*}
\end{proof}

\section{Invasion reproduction number (IRN)}\label{app_sec:IRN_Imperfect}
 The infectious classes $E_{i}, I_{i} , ~\&Q_{i}$ are linearized when strain-$c$ is resident. The vector of new infections $\mathcal{F}$ and outflow vector $\mathcal{V}$ are given by
\begin{equation}
\begin{array}{lll}
	\mathcal{F}=
\begin{pmatrix}
\xi_{i}S + \eta_cR_c\xi_i\\
0\\
0\\
\end{pmatrix}& and &\mathcal{V}=
\begin{pmatrix}
	\mathbf{p}_iE_i\\
\mathbf{q}_iI_i -\rho_iE_i+\tau_iQ_i\\
\mathbf{r}_iQ_i-\alpha_iE_i-\sigma_iI_i\\
\end{pmatrix}\\
\end{array}
\end{equation}
We take the derivatives of $\mathcal{F}$ and $\mathcal{V}$ and evaluate at \cref{eqn:Covid_Endemic_Strain_Imperfect} to obtain matrices, namely:

\begin{equation}
	\begin{array}{lll}
	F=
		\begin{pmatrix}
0&\tilde{S_c^*}\beta_i +\tilde{R_c^*}\beta_i\eta_ c&0\\	
0&0&0\\
0&0&0\\
		\end{pmatrix}& and &V=
		\begin{pmatrix}
		\mathbf{p}_{i}&0&0\\	
		-\rho_i&	\mathbf{q}_{i}&-\tau_{i}\\
		-\alpha_i&-\sigma_i&	\mathbf{r}_{i}\\
		\end{pmatrix}\\
	\end{array}
\end{equation}

The strain-$i$ invasion reproduction number is then defined as $_i\tilde{\mathcal{R}}_c^q= \rho(FV^{-1})$.

\section*{Acknowledgments}
A.D.F. thanks the German Academic Exchange Service  for the doctoral studies financial supports.

\bibliographystyle{siamplain}
\bibliography{references}
\end{document}